\documentclass[12pt]{article}
\pdfoutput=1

\usepackage[pdftex]{graphicx}
\usepackage{amssymb}
\usepackage{amsmath}
\usepackage{cite}
\usepackage{multirow}
\usepackage{gensymb}
\usepackage{longtable}
\usepackage{booktabs}
\usepackage{array}
\usepackage{anyfontsize} 
\usepackage{ulem}
\usepackage{wrapfig}
\usepackage{xcolor}
\usepackage{cancel}
\usepackage{arydshln}

\definecolor{tit}{rgb}{0.1,0.2,0.4}

\newcommand{\eq}[1]{\begin{equation} #1 \end{equation}}

\newcommand{\eqa}[1]{\begin{eqnarray} #1 \end{eqnarray}}

\newcommand{\GeV}{\,{\rm GeV}}
\newcommand{\MeV}{\,{\rm MeV}}

\newcommand{\op}{\mathcal{O}}

\newcommand{\Ref}[1]{Ref.~\cite{#1}}
\newcommand{\Eq}[1]{Eq.~(\ref{#1})}

\newcommand{\bea}{\begin{eqnarray}}
\newcommand{\eea}{\end{eqnarray}}
\newcommand{\be}{\begin{equation}}
\newcommand{\ee}{\end{equation}}
\newcommand{\non}{\nonumber\\ }
\newcommand{\vsl}{ v \hspace{-2.2truemm}/ }

\newcommand{\xsl}{ x \hspace{-2.2truemm}/ }

\newcommand{\im}{{\rm Im}}

\definecolor{DRed}{rgb}{0.8,0,0.1}
\definecolor{DBlue}{rgb}{0,0,0.8}

\usepackage[titles]{tocloft}

\begin{document}

% \allowdisplaybreaks

\begin{flushright}
{\small
SI-HEP-2016-09\\
QFET-2016-04\\
NIOBE-2017-01
%{\color{red} DRAFT of \today }
}
\end{flushright}
$\ $
\vspace{-2mm}

\begin{center}
\fontsize{18.2}{20}\selectfont
\bf
\color{tit}
$B\to\pi\pi$ Form Factors from Light-Cone Sum Rules\\[2mm] with $B$-meson Distribution Amplitudes
\end{center}

\vspace{1mm}
\begin{center}
%\large
{\sc Shan Cheng}$^{\,a}$, {\sc Alexander Khodjamirian}$^{\,a}$  and {\sc Javier Virto}$^{\,b}$\\[5mm]
{\it \small
$^a$ Theoretische Physik 1, Naturwissenschaftlich-Technische Fakult\"at,\\[1mm]
Universit\"at Siegen, 57068 Siegen, Germany
\\[3mm]
$^b$ Albert Einstein Center for Fundamental Physics, Institute for Theoretical Physics,\\
University of Bern, CH-3012 Bern, Switzerland.
}
\end{center}

\vspace{1mm}
\begin{abstract}\noindent
\vspace{-5mm}

We study $B\to\pi\pi$ form factors using QCD light-cone sum rules with $B$-meson distribution amplitudes.
These form factors describe the semileptonic decay \mbox{$B\to \pi\pi \ell\bar{\nu}_{\ell}$}, and constitute an essential input
in $B\to \pi\pi \ell^+\ell^-$ and $B\to \pi\pi\pi$ decays.
We  employ the correlation functions where a dipion isospin-one state is interpolated by the vector light-quark current.
We obtain sum rules where convolutions of the $P$-wave $\bar{B}^0\to \pi^+\pi^0$ form factors with the timelike pion vector
form factor are related to universal $B$-meson distribution amplitudes.
These sum rules are valid in the kinematic regime where the dipion state has a large energy and a low invariant mass,
and reproduce analytically the known light-cone sum rules for $B\to \rho$ form factors in the limit of $\rho$-dominance
and zero width, thus providing a systematics for so far unaccounted corrections to $B\to\rho$ transitions.
Using data for the pion vector form factor, we estimate finite-width effects and the contribution of excited $\rho$-resonances
to the $B\to\pi\pi$ form factors.
We find that these contributions amount up to $\sim 20\%$ in the small dipion  mass region 
where they can be effectively regarded as a nonresonant ($P$-wave) background to the $B\to\rho$ transition.

\end{abstract}

\newpage

\tableofcontents

\bigskip

\section{Introduction}\label{introduction}

The $B\to \pi\pi$ transition form factors encode the rich hadronic dynamics accompanying the short-distance $ b\to u$ transition in
the semileptonic $B\to \pi\pi \ell\nu_\ell$ ($B_{\ell 4}$) decays (see e.g.~\cite{Faller:2013dwa,Kang:2013jaa}), which may provide a
competitive determination of the CKM parameter $|V_{ub}|$~\cite{Sibidanov:2013rkk} if an accurate knowledge of the form factors can
be assessed. The $B\to \pi\pi $ form factors are also an essential hadronic input to the rare flavor-changing neutral-current
decay $B\to \pi\pi \ell^+\ell^-$  \cite{Aaij:2014lba} and to nonleptonic three-body  $B$ decays such as
$B\to \pi\pi\pi$ \cite{Krankl:2015fha,1609.07430}.

While $B\to \pi\pi$ form factors are dominated by the resonant $B\to \rho$ transition (which has been studied extensively in the narrow-width approximation),
finite-width effects and ``nonresonant" contributions have not yet been addressed systematically.
These effects are considerably more difficult to describe theoretically, providing non-trivial challenges for both analytical methods
and lattice simulations. At large dipion invariant masses, the form factors can be calculated in QCD
factorization~\cite{BoerFeldmannDyk}. For small dipion masses at low hadronic recoil, heavy-meson chiral perturbation
theory may be combined with dispersion theory as proposed in \Ref{Kang:2013jaa}. At large hadronic recoil (and low dipion
invariant mass), the method of light-cone sum rules (LCSRs) is operative, and has been used in \Ref{Hambrock:2015aor} in
terms of dipion distribution amplitudes (DAs) \cite{Diehl:1998dk,Polyakov:1999ZE}. However, the limited knowledge of these
DAs asks for other QCD based  methods to access the $B\to \pi\pi$ form factors in the same kinematic regime.

In this paper we propose to use the LCSRs with $B$-meson DAs \cite{Khodjamirian:2005ea,Khodjamirian:2006st,DeFazio:2005dx}.
For definiteness we will focus on the transition $\bar{B}^0\to \pi^+ \pi^0$ with the isospin-one final dipion state;
in the future these sum rules can be easily extended to the other isospin states.
We will obtain a set of sum rules where the hadronic representation contains the $B\to \pi\pi$ form factors of
interest convoluted with the timelike pion vector form factor. The latter is very accurately measured 
within a wide range of dipion masses. The sum rules 
obtained in this paper reproduce the known  sum rules for the $B\to \rho$ form factors in the limit of $\rho$-dominance
and zero width, and can be used to test models for $B\to \pi\pi$ form factors. We will illustrate this point by performing
a numerical study of the effects of excited $\rho$-resonances within a three-resonance model that fits the pion form
factor accurately, assessing the deviations from $\rho$-meson dominance in $B\to \rho$ transitions.

The plan of the paper is the following. In Section~\ref{lcsrs} we derive the LCSRs with $B$-meson DAs for the  full set of vector and axial-vector $B\to\pi\pi$ form factors. The current accuracy of the operator-product expansion (OPE) includes the contributions of two- and three-particle DAs. In Section~\ref{res-model} we adopt a model for $B\to \pi\pi$ form factors in terms of $\rho$-resonances.
The LCSRs are then rewritten in the form of relations containing the model parameters. These relations are analyzed numerically in Section~\ref{numerics}, taking as input a similar
model for the pion vector form factor, fitted to the experimental data on $\tau\to \pi^+\pi^0\nu_\tau$.  This analysis will allow us to quantify the deviations from the $\rho$ dominance approximation. We conclude in Section~\ref{conclusion}.
The appendices contain: (\ref{app:B-DA}) the relevant formulae for $B$-meson DAs used in LCSRs, (\ref{app:pi-ff}) the model for the pion timelike form factor, and (\ref{app:2pt-s0}) the two-point sum rule used to fix the effective threshold in the sum rules.

\section{Light-Cone Sum Rules}\label{lcsrs}

\begin{figure}
\centering
\includegraphics[width=0.7\textwidth]{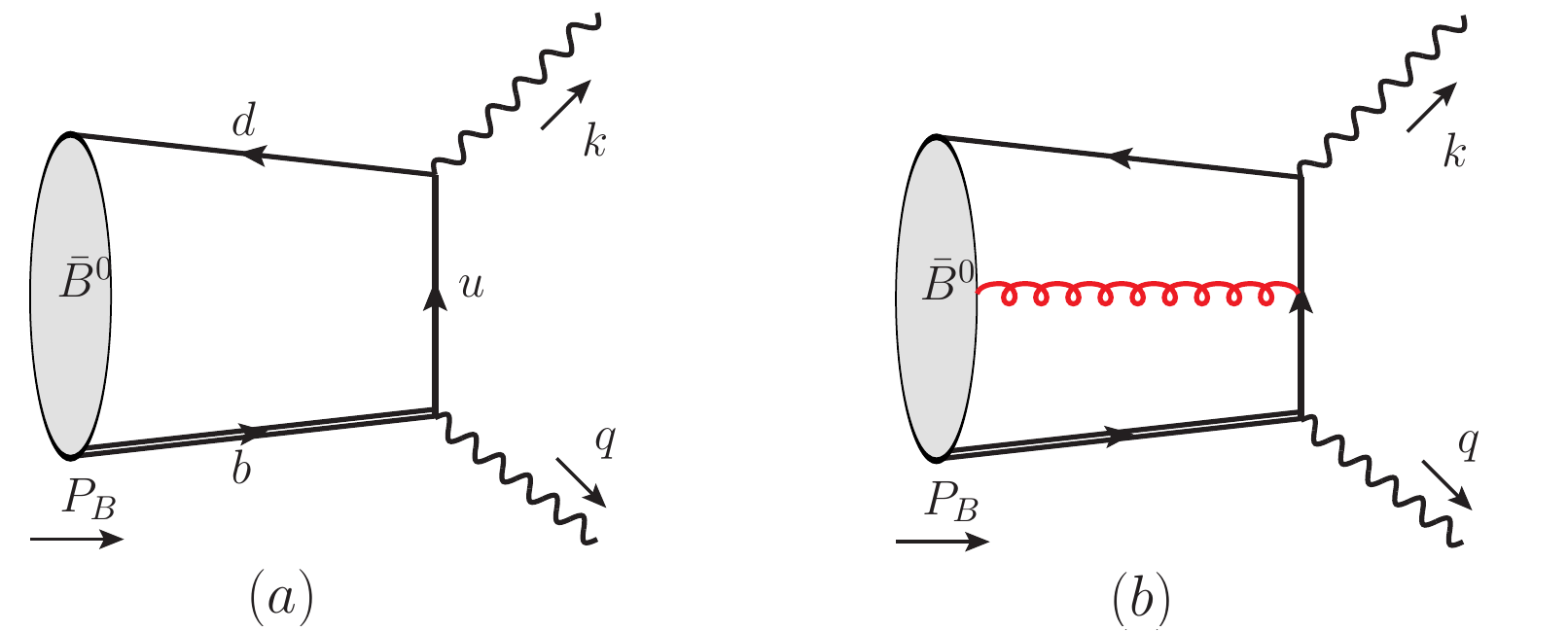}
\caption{\it Diagram of the correlation function (\ref{eq:corr}): (a) at leading order (two-body B-meson DA), (b) the soft-gluon contribution (three-body B-meson DA). Wavy lines with four-momentum $k$ and $q$ represent the dipion interpolating and weak $b\to u$ currents, respectively.}
\label{fig:fig1}
\end{figure}

Following \Ref{Khodjamirian:2006st} we introduce the correlation function of the $\bar{d}\gamma_\mu u$ interpolation current with the $b\to u$ weak current:
\be
F_{\mu\nu}(k,q)=i\int d^4x e^{i k \cdot x} \langle 0 | {\rm T} \{\bar{d}(x) \gamma_\mu u(x),
\bar{u}(0)\gamma_\nu (1-\gamma_5) b(0) \} | \bar{B}^0(q+k) \rangle,
\label{eq:corr}
\ee
sandwiched between the on-shell $B$-meson and vacuum states. The four-momenta of the currents are $k$ and $q$ respectively, so that $(q+k)^2=m_B^2$.
The correlation function (\ref{eq:corr}) is decomposed into independent Lorentz structures:
\bea
F_{\mu\nu}(k,q)=
\varepsilon_{\mu\nu\rho\sigma}q^{\rho}k^{\sigma}F_{(\varepsilon)}(k^2,q^2)
+i g_{\mu\nu}F_{(g)}(k^2,q^2) +iq_\mu k_\nu F_{(qk)}(k^2,q^2)\non
+ik_\mu k_\nu F_{(kk)}(k^2,q^2)+iq_\mu q_\nu F_{(qq)}(k^2,q^2) +ik_\mu q_\nu F_{(kq)}(k^2,q^2)\,,
\label{eq:corr_expan}
\eea
where the first term\,\footnote{
In this paper we use the conventions $\ \varepsilon_{0123} = - \varepsilon^{0123} = +1$ and $\gamma_5 \equiv (i/4!)\,\varepsilon^{\mu\nu\rho\sigma} \gamma_\mu \gamma_\nu \gamma_\rho \gamma_\sigma$.
} corresponds to the contribution of the vector $b\to u$ current. Only the structures in the first line will be used in the sum rules below.

In the region $q^2\ll m_B^2$  and $|k^2|\gg \Lambda_{QCD}^2$, due to the large virtuality of the intermediate $u$-quark, the correlation function is calculable by means of an OPE, involving the DAs of the $B$-meson defined in HQET. To leading order, one contracts the $u$-quark fields in (\ref{eq:corr}) as a free propagator, neglecting hereafter the $u$-quark mass. The remaining  heavy-light bilocal quark-antiquark operators sandwiched between the $B$-meson and vacuum states are parametrized by the two-particle DAs $\phi^B_{+,-}(\omega)$, where $\omega$ is related to the momentum of the light-quark in the $B$ meson. We will also include  the corrections due to a low virtuality (``soft'') gluon emitted from the propagator and absorbed in the three-particle $B$-meson DAs \cite{Khodjamirian:2006st}. The diagrams corresponding to the two contributions to the correlation function are shown in Fig.~\ref{fig:fig1}. The definitions of the DAs are listed in Appendix \ref{app:B-DA} together with the models we will use to describe them.

To outline the derivation of the sum rule, we choose the invariant amplitude $F_{(\varepsilon)}(k^2,q^2)$, for which the corresponding OPE result can be written as
\bea
F_{(\varepsilon)}^{\rm OPE}(k^2,q^2) =
f_B m_B \ \int_0^{\infty} d\sigma\ \frac{\phi_+^B(\sigma m_B)}{\bar\sigma(s-k^2)}+ \cdots \ ,
\label{eq:OPE-2}
\eea
where
\be
s=s(\sigma,q^2)=\sigma m_B^2 - \sigma q^2/\bar{\sigma}\ , \qquad \bar{\sigma}\equiv 1-\sigma \ ,
\label{eq:s_sig}
\ee
and the ellipsis denotes the subleading 3-particle DA contributions calculated in \Ref{Khodjamirian:2006st}. The OPE expression (\ref{eq:OPE-2}) has the form of a dispersion integral in the variable $k^2$:
\be
F^{\rm OPE}_{(\varepsilon)}(k^2,q^2)= \frac{1}{\pi}\
\int_0^\infty ds\ \frac{\textmd{Im} F_{(\varepsilon)}^{\rm OPE}(s,q^2)}{s-k^2}\,,
\label{eq:disp}
\ee
with the imaginary part given by
\be
\frac{1}{\pi}\,\im F_{(\varepsilon)}^{\rm OPE}(s,q^2)= f_B m_B \
\Bigg[\left(\frac{d\sigma}{ds}\right)\frac{\phi_+^B(\sigma m_B)}{(1-\sigma)}\Bigg]_{\sigma(s)}+ \cdots \,,
\label{eq:imOPE}
\ee
where $\sigma(s)$ is obtained by solving Eq.~(\ref{eq:s_sig}).

In parallel, for the same invariant amplitude we employ the hadronic dispersion relation in the variable $k^2$,
\be
F_{(\varepsilon)}(k^2,q^2)= \frac{1}{\pi}\
\int\limits_{4m_\pi^2}^\infty ds\ \frac{\textmd{Im} F_{(\varepsilon)}(s,q^2)}{s-k^2}\,.
\label{eq:had_disp}
\ee
The hadronic spectral function of the correlation function is obtained from the unitarity relation, that is, inserting the complete set of states with quantum numbers of the $\bar{d}\gamma_\mu u$ current between the two currents in \Eq{eq:corr}:
\eq{
2\, \im F_{\mu\nu}(k,q)= \int d\tau_{2\pi}
\langle 0 |\bar{d} \gamma_\mu u \,| \pi^+(k_1)\pi^0(k_2) \rangle
\langle \pi^+(k_1)\pi^0(k_2) | \bar{u} \gamma_\nu (1-\gamma_5) b | \bar{B}^0(q+k)\rangle
+ \cdots\,,
\label{eq:unit}
}
where the lowest intermediate dipion state is included explicitly and the ellipsis denotes the contributions from other intermediate states with higher thresholds: $4\pi, K\bar{K}$, etc. This hadronic representation is more general than the single-pole approximation adopted in Ref.~\cite{Khodjamirian:2006st}, where the two-pion-state contribution was replaced by a single narrow  $\rho$-meson state.

We use the definition of the pion vector form factor:
\be
\langle \pi^+(k_1)\pi^0(k_2) |\bar{u}\gamma_\mu d | 0 \rangle = -\sqrt{2}\ \overline k_\mu\, F_{\pi}(k^2),
\label{eq:piFF}
\ee
where $k = k_1+k_2$ and $\bar{k}=k_1-k_2$. In the isospin symmetry limit $F_{\pi}(k^2) = F^{\rm em}_\pi(k^2)$, where  the pion electromagnetic form factor is normalized as $F^{\rm em}_\pi(0)=1$. We also adopt the following definition for the $B\to \pi\pi$ form factors:\,\footnote{
See e.g. Ref.~\cite{Faller:2013dwa}. Here we use the phase convention of Ref.~\cite{Hambrock:2015aor}.
}
\begin{eqnarray}
&&\hspace{-1cm}
i\langle \pi^+(k_1) \pi^0(k_2)|\bar{u}\gamma_\nu(1-\gamma_5)b|\bar{B}^0(p)\rangle
= F_\perp (k^2,q^2,q\cdot \overline{k})\, \frac{2}{\sqrt{k^2} \sqrt{ \lambda}} \,
i\epsilon_{\nu\alpha\beta\gamma} \, q^\alpha \, k^{\beta} \, \bar{k}^{\gamma}
\nonumber \\
&&\hspace{2cm}
+ F_t (k^2,q^2,q\cdot \overline{k})\, \frac{q_\nu}{\sqrt{q^2}}
+ F_0 (k^2,q^2,q\cdot \overline{k})\, \frac{2\sqrt{q^2}}{\sqrt{\lambda}} \,
\Big(k_\nu - \frac{k \cdot q}{q^2} q_\nu\Big)
\nonumber\\
&&\hspace{2cm}
+ F_\parallel(k^2,q^2,q\cdot \overline{k}) \, \frac{1}{\sqrt{k^2}} \,
\Big(\overline{k}_\nu - \frac{4 (q\cdot k) (q \cdot \overline{k})}{\lambda} \, k_\nu
+ \frac{4 k^2 (q\cdot \overline{k})}{\lambda} \, q_\nu \Big)\,,
\label{eq:formf}
\end{eqnarray}
where $\lambda\equiv\lambda(m_B^2,q^2,k^2) = m_B^4 + q^4 + k^4 - 2(m_B^2 q^2 + m_B^2 k^2 + q^2 k^2 )$ is the kinematic K\"all\'en function. In addition, $q\cdot k=\frac12(m_B^2-q^2-k^2)$ and
\be
q\cdot \overline k =\frac12 \sqrt{\lambda}  \ \beta_\pi(k^2) \cos\theta_\pi\,,
\label{eq:theta}
\ee
where $\beta_\pi(k^2)= \sqrt{1-4m_\pi^2/k^2}$, and $\theta_\pi$ is the angle between the 3-momenta of the neutral pion and the $B$-meson in the dipion rest frame. Note that the form factor $F_\perp$ in the decomposition of Eq.~(\ref{eq:formf}) parametrizes the transition  matrix element of the vector weak $b\to u$ current, whereas the other three  form factors correspond to the axial-vector weak current.

For the form factors in Eq.~(\ref{eq:formf}) we will use the partial wave expansions
\bea
F_{0,t}(k^2,q^2, q\cdot \bar k)
&=&\sqrt{3}\,F_{0,t}^{(\ell=1)}(k^2,q^2)\,P_1^{(0)}(\cos\theta_\pi) + \cdots\,,
\label{eq:ffexpansionhel1}
\non
F_{\perp,\parallel}(k^2,q^2,q\cdot \bar k)
&=&\sqrt{3}\,F_{\perp,\parallel}^{(\ell=1)}(k^2,q^2)\,
\frac{P_1^{(1)}(\cos\theta_\pi)}{\sin \theta_\pi}+ \cdots\,,
\label{eq:ffexpansionhel2}
\eea
where $P_1^{(0)}(\cos\theta_\pi)=\cos \theta_\pi$ and $P_1^{(1)}(\cos\theta_\pi)=-\sin\theta_\pi$ are the associated Legendre polynomials.
Only the $P$-wave $(\ell=1)$ components shown explicitly in the above expansions survive in the convolution of the two hadronic matrix elements in Eq.~(\ref{eq:unit}). Indeed, the hadronic matrix element of the local $J^P=1^-$ current $\bar{d}\gamma_\mu u$ parametrized with the pion vector form factor contains only the $P$-wave dipion contribution and hence effectively serves as a $P$-wave projector for the $B\to \pi \pi$ form factors. In order to extend the method suggested here to other partial waves one needs to replace the $\bar{d}\gamma_\mu u$ interpolating current in the correlation function with a different current or combination of currents.

Substituting the definitions (\ref{eq:piFF}) and (\ref{eq:formf}) in Eq.(\ref{eq:unit}), integrating over the angles in the dipion phase space and sorting out the different kinematic structures, we obtain the imaginary parts of all relevant invariant amplitudes. In particular, the one generated by the vector $b\to u$ current reads:
\be
\frac{1}{\pi} \ \im F_{(\varepsilon)}(s,q^2)=
\frac{\sqrt{s}\,[\beta_{\pi}(s)]^3}{4\sqrt{6}\pi^2 \sqrt{\lambda} }
\,F^\star_{\pi}(s) \, F_{\perp}^{(\ell=1)}(s,q^2) + \cdots \ ,
\label{eq:imF}
\ee
where hereafter $\lambda\equiv\lambda(m_B^2,q^2,s)$ and again the  ellipsis denote contributions from the intermediate states $4\pi, \bar K K$, etc. Judging by studies on pion form factors at $s \lesssim 1.0\,\text{-}\, 1.5\ \GeV^2$, these contributions are expected to be suppressed  (see e.g. \Ref{Eidelman:2003uh} and the discussion in \Ref{Kang:2013jaa}).

We then insert the hadronic spectral function (\ref{eq:imF}) in the r.h.s. of \Eq{eq:had_disp}. For the l.h.s. we use \Eq{eq:disp}, as the OPE is a good approximation to the correlation function in the region $-k^2\gg \Lambda_{\rm QCD}^2$. At this point, we Borel-transform both sides of the resulting equality,  effectively replacing  the variable $k^2$ with the Borel parameter squared $M^2$.  In addition, we employ the quark-hadron duality approximation, which amounts to the assumption that the integrals over the hadronic spectral density $ \im F_{(\varepsilon)}(s)$ and over $ \im F_{(\varepsilon)}^{\rm OPE}$ are equal:
\be
\int\limits_{s_0^{2\pi}}^\infty ds \ e^{-s/M^2} \,\im F_{(\varepsilon)}(s,q^2)=
\int\limits_{s_0^{2\pi}}^\infty ds \ e^{-s/M^2} \,\im F^{\rm OPE}_{(\varepsilon)}(s,q^2)\,,
\label{eq:dual}
\ee
where  $s_0^{2\pi}$ is the effective threshold. The above semi-local duality relation  allows one to effectively cut-off the integrals over the dipion mass in the sum rule.  Note that we use a quark-hadron duality ansatz,  which is more general than a local duality, that would assume equality of the integrands on both sides  of Eq.~(\ref{eq:dual}) for every $s>s_0^{2\pi}$. Note also that the falling Borel exponent (provided $M^2$ is not too large) suppresses the large $s$ region of the integrals, making the duality relation less sensitive to multihadron states with thresholds larger than $s_0^{2\pi}$. Depending on the choice of $s_0^{2\pi}$, the $4 \pi$ and $K\bar{K}$ states may still contribute to the region $4m_\pi^2<s < s_0^{2\pi}$ but their expected suppression with respect to the dipion state justifies to retain only the latter in $\textmd{Im} F_{(\varepsilon)}(s,q^2)$.

Finally, we obtain the following LCSR:
\bea
&&\hspace{-25mm} \int_{4m_{\pi}^2}^{s_0^{2\pi}} ds~e^{-s/M^2} \frac{\sqrt{s}\,[\beta_{\pi}(s)]^3}{4\sqrt{6}\pi^2\sqrt{\lambda}}\,F^\star_{\pi}(s) \,F_{\perp}^{(\ell=1)}(s,q^2)\non[2mm]
\hspace{10mm}&=& f_B m_B\ \Bigg[\int_0^{\sigma_0^{2\pi}} d\sigma~e^{-s(\sigma,q^2)/M^2}\
\frac{\phi_+^B(\sigma m_B)}{\bar \sigma} + m_B\,\Delta V^{BV}(q^2,\sigma_0^{2\pi},M^2)
\Bigg]\,,
\label{eq:lcsrFperp}
\eea
where $\sigma_0^{2\pi}$ is the solution of the relation $\sigma m_B^2-\sigma q^2/\bar{\sigma}=s^{2\pi}_0$ and the
explicit expression for the three-particle DA contribution $\Delta V^{BV}$ can be found in the Appendix of
Ref.~\cite{Khodjamirian:2006st}.\footnote{
Note that the factor $e^{m_V^2/M^2}$ has to be removed from the integrand.
}

Repeating the same steps for the invariant amplitudes $F_{(g)}$ and $F_{(qk)}$ in Eq.~(\ref{eq:corr_expan}),
we obtain two additional LCSRs containing the integrals over the $B\to \pi\pi$ form factors $F_{\parallel,0}^{(\ell =1)}$
and $F_{0}^{(\ell=1)}$ respectively:
\bea
&& \hspace{-12mm} \int_{4m_{\pi}^2}^{s^{2\pi}_0} ds\ e^{-s/M^2}
\frac{\sqrt{s}\,[\beta_{\pi}(s)]^3}{4\sqrt{6}\pi^2} \,F^\star_{\pi}(s) \,F^{(\ell=1)}_{\parallel}(s,q^2)\non[2mm]
&=&f_B m_B\ \Bigg[ \int_0^{\sigma^{2\pi}_0}  d\sigma \ e^{-s(\sigma,q^2)/M^2}\,
\frac{\bar{\sigma}^2m_B^2-q^2}{\bar{\sigma}^2} \phi_+^B(\sigma m_B)
+ m_B^2 \,\Delta A_1^{BV}(q^2,\sigma_0^{2\pi},M^2)
\Bigg]\,,\ 
\label{eq:LCSRs-AV-g}
\eea
and
\bea
&&\hspace{-7mm} \int_{4m_{\pi}^2}^{s^{2\pi}_0} ds \ e^{-s/M^2}
\frac{\sqrt{s}\,[\beta_{\pi}(s)]^3}{2 \sqrt{6} \pi^2 \lambda} \,F^\star_{\pi}(s)
\Bigg[\frac{ (m_B^2-q^2-s)}{2}  F^{(\ell=1)}_{\parallel}(s,q^2)
+ \frac{\sqrt{s}\sqrt{q^2}}{\beta_{\pi}(s)} F^{(\ell=1)}_{0}(s,q^2)\Bigg]
\nonumber\\[2mm]
&=&f_B m_B \ \Bigg\{ \int_0^{\sigma^{2\pi}_0} \!\! d\sigma\ e^{-s(\sigma,q^2)/M^2}
\ \bigg[
\frac{\bar{\sigma}-\sigma}{\bar{\sigma}}
\phi_+^B(\sigma m_B)
+\frac{2\sigma \bar{\sigma}m_B^2}{\bar{\sigma}^2m_B^2-q^2}
\Big[\phi_+^B(\sigma m_B) - \phi_-^B(\sigma m_B) \Big]
\nonumber\\[2mm]
&&\hspace{13mm} + \left(\frac{4\sigma \bar \sigma^{2} m_B^3}{(\bar{\sigma}^2m_B^2-q^2)^2} + \frac{2(1-2\sigma)m_B}{\bar{\sigma}^2m_B^2-q^2}\right)\bar{\Phi}_{\pm}^B(\sigma m_B) \bigg ]
+ \Delta A_2^{BV}(q^2,\sigma_0^{2\pi},M^2)
\Bigg\}\,,
\label{eq:LCSRs-AV-qp}
\eea
where  $\bar{\Phi}_{\pm}^B$ is defined in Appendix~\ref{app:B-DA}, and the three-particle  contributions $\Delta A_{1,2}^{BV}(q^2,\sigma_0^{2\pi},M^2)$
can be again found in the Appendix of Ref.~\cite{Khodjamirian:2006st}.$^\arabic{footnote}$

The remaining Lorentz structures in the correlation function provide additional, more complicated relations  between the three form factors, hence we do not consider them here. Instead, we obtain a new sum rule for the ``timelike-helicity'' form factor $F^{(1)}_{t}$ by considering a different correlation function with the pseudoscalar heavy-light current,
\bea
F_{\mu}(k,q)&=& i\int d^4x e^{i k \cdot x} \langle 0 | {\rm T} \{\bar{d}(x) \gamma_\mu u(x),
im_b\bar{u}(0)\gamma_5 b(0) \} | \bar{B}^0(q+k) \rangle\nonumber\\[2mm]
&=&  i k_\mu F_{(k)}(k^2,q^2) + i q_\mu F_{(q)}(k^2,q^2).
\label{eq:corr-PS}
\eea
The form factor $F_{t}$ can be isolated by multiplying both sides of Eq.~(\ref{eq:formf}) with $q_\mu$, giving
\be
\langle \pi^+(k_1) \pi^0(k_2)|\, im_b\bar{u} \gamma_5 b\, |\bar{B}^0(p)\rangle
= \sqrt{q^2}\ F_{t}(k^2,q^2,q\cdot \overline{k})\,.
\label{eq:formf-ps}
\ee
After inserting the dipion intermediate state in \Eq{eq:corr-PS}, only the above form factor contributes. Considering the invariant amplitude $F_{(q)}$ and carrying out a similar derivation as for the previous correlation function, we obtain the following LCSR for $F^{(\ell=1)}_{t}$:
\bea
&&\hspace{-5mm} \int_{4m_{\pi}^2}^{s_0^{2\pi}} ds~e^{-s/M^2}~
\frac{s\ \sqrt{q^2}\ [\beta_{\pi}(s)]^2 }{4 \sqrt{6} \pi^2 \sqrt{\lambda}}
\,F^\star_{\pi}(s)\,F_{t}^{(\ell=1)}(s,q^2)
\label{eq:lcsrFtrans} \\[2mm]
&=& - f_B m_B^2 m_b\ \Bigg\{ \int_0^{\sigma^{2\pi}_0} d\sigma~e^{-s(\sigma,q^2)/M^2}~
\Bigg[\frac{\sigma}{\bar{\sigma}} \phi_-^B(\sigma m_B)
-\frac{\bar{\Phi}_{\pm}^B(\sigma m_B)}{\bar{\sigma}m_B} \Bigg]
+ \Delta A_0^{BV}(q^2,\sigma_0^{2\pi},M^2)
\Bigg\}\,,\nonumber
\eea
where the OPE result in the r.h.s. is new and has not been given before.
The new expression for the three-particle contribution  $\Delta A_0^{BV}(q^2,\sigma_0^{2\pi},M^2)$  is given explicitly in  Appendix~\ref{app:B-DA}.

The sum rules in Eqs.~(\ref{eq:lcsrFperp})-(\ref{eq:LCSRs-AV-qp}) and (\ref{eq:lcsrFtrans})  with generalized hadronic part represent the main results of this paper. They provide additional constraints and normalization for $B\to \pi\pi$ form factors if  one adopts a certain ansatz or model for them. On the other hand, if the $B\to \pi\pi$ form factors are calculated via an alternative method, one can check the validity and consistency of the results. In addition to the universal $B$-meson DAs, the pion vector form factor in the timelike region represents the necessary input in these sum rules. The magnitude of this form factor is well known experimentally from $\tau^- \to \pi^-\pi^0 \nu_\tau$~\cite{Fujikawa:2008ma} and $e^+e^-\to \pi^+\pi^-$~\cite{Aubert:2009AD}.

Our last comment in this section concerns the final state interaction phase. Below the inelastic threshold for the pion form factor this phase coincides with the dipion elastic scattering phase according to Watson's theorem. Here an analogous condition should be fulfilled  in the adopted approximation of two-pion intermediate state in the hadronic dispersion relation. Due to the reality of the imaginary parts, such as the one in  Eq.~(\ref{eq:imF}),  the strong phase of all $B\to \pi\pi$ form factors should be universal (modulo $\pi$) and equal to the phase of the pion vector form factor:
\be
\im \Big[F_k^{(\ell=1)}(s,q^2) \  F^\star_\pi(s)\Big] =0 \ ,
\quad k= \{\perp,\parallel,0,t\}\ .
\label{eq:phaseeq}
\ee
This condition was already mentioned and used in the elastic scattering region  in Ref.~\cite{Kang:2013jaa}.

Note that since Eq.~(\ref{eq:phaseeq})  follows from the
general unitarity relation, it enforces any parametrization of 
the $B\to \pi \pi$ form factors to be chosen such that
the phases of separate $q^2$-dependent components in each
form factor can only depend on $s$, being  correlated with the
phase of the pion form factor. In the following we will take this
condition into account when choosing a particular
ansatz for the $B\to \pi\pi$ form factors.

\section{Probing resonance models for $B\to \pi\pi$ form factors}
\label{res-model}

Originally, the LCSRs with $B$-meson DAs derived from the correlation function in \Eq{eq:corr} were used in
Ref.~\cite{Khodjamirian:2006st} to determine the $B\to \rho$ form factors. In what follows we use the standard
definition of these form factors:
\bea
\langle \rho^+(k) |\bar{u}\gamma_\nu(1-\gamma_5) b |\bar{B}^0(p)\rangle=
\epsilon_{\nu\alpha\beta\gamma}
\epsilon^{* \alpha} q^\beta k^\gamma\frac{2V^{B\rho}(q^2)}{m_B+m_\rho}
%\nonumber\\
-i\epsilon^{*}_\nu (m_B+m_\rho)A_1^{B\rho}(q^2)
\nonumber\\
+
i(2k+q)_\nu(\epsilon^{*}\cdot q)\frac{A_2^{B\rho}(q^2)}{m_B+m_\rho}+
iq_\nu(\epsilon^{*}\cdot q)\frac{2m_\rho}{q^2}\big(
A_3^{B\rho}(q^2)-A_0^{B\rho}(q^2)\big)\,,
\label{eq:BrhoFFdef}
\eea
where $2m_\rho\, A_3^{B\rho}(q^2) =(m_B+m_\rho)A_1^{B\rho}(q^2)-(m_B-m_\rho)A_2^{B\rho}(q^2)$.

To recover the sum rules obtained in \Ref{Khodjamirian:2006st} from the sum rules derived in the previous section, one has to employ the dispersion relation in $k^2=s$ for the $B\to\pi\pi$ form factors retaining only the single $\rho$-pole contribution. E.g., for the vector-current form factor one has:
\be
\frac{\sqrt{3}F_{\perp}^{(\ell=1)}(s,q^2)}{\sqrt{s}\sqrt{\lambda}}=
\frac{g_{\rho\pi\pi}V^{B\rho}(q^2)}{(m_B+m_\rho)\big[m_\rho^2-s-i\sqrt{s}\ \Gamma_\rho(s)\big]} +\cdots
\label{eq:VrelFperp}
\ee
where the excited state contributions with the  $\rho$ quantum numbers indicated by the ellipsis are assumed to be  accounted for by the duality approximation. Note that in the above, for the sake of generality, we go beyond the narrow $\rho$ approximation and adopt the energy-dependent $\rho\to \pi\pi$ width
\be
\Gamma_\rho(s)=
\frac{g_{\rho\pi\pi}^2[\beta_\pi(s)]^3\sqrt{s}}{48 \pi} \theta(s-4 m_\pi^2)
=\Gamma^{\rm tot}_\rho \left[\frac{\beta_\pi(s)}{\beta_\pi(m_\rho^2)}
\right]^3\frac{\sqrt{s}}{m_\rho}\theta(s-4 m_\pi^2)\,,
\label{eq:width}
\ee
so that $\Gamma^{\rm tot}_\rho$ is the total width and  the function $\Gamma_\rho(s)$ vanishes below the dipion threshold $s=4m_\pi^2$. The energy-dependent width can be interpreted as a result of the resummation of two-pion loops coupled to the $\rho$ state. For consistency,  a $\rho$-dominance approximation for the pion form factor $F_\pi$ has to be adopted too:
\be
F^\star_{\pi}(s)=\frac{f_\rho g_{\rho\pi\pi}m_\rho}{\sqrt{2}(m_\rho^2-s+i\sqrt{s}\ \Gamma_\rho(s))}\,,
\label{eq:VrelFpi}
\ee
where the $\rho$-meson  decay constant and strong coupling are normalized as:
\be
\langle \rho^+|\bar{d} \gamma_\mu u |0\rangle=f_\rho m_\rho \epsilon^\star_\mu, ~~
\langle \pi^+(k_1)\pi^0(k_2)|\rho^+\rangle= g_{\rho\pi\pi}(k_1-k_2)^\alpha
\epsilon_\alpha\ .
\label{eq:frogrho}
\ee
Using the approximations of Eqs.~(\ref{eq:VrelFperp}) and~(\ref{eq:VrelFpi}) and taking into account Eq.~(\ref{eq:width}), the l.h.s. of Eq.~(\ref{eq:lcsrFperp}) becomes
\be
\frac{2 f_\rho m_\rho V^{B\rho}(q^2) }{(m_B+m_\rho)}
\int_{4m_{\pi}^2}^{s_0^{2\pi}} ds~e^{-s/M^2}\Bigg(
\frac 1{\pi}\frac{\Gamma_\rho(s)\sqrt{s}}{(m_\rho^2-s)^2+s\Gamma^2_\rho(s)}
\Bigg)
\xrightarrow{\Gamma_\rho^{\rm tot}\to 0} \,\,\frac{2 f_\rho m_\rho V^{B\rho}(q^2) }{(m_B+m_\rho)}e^{-m_\rho^2/M^2}\,,
\label{eq:rhodomin}
\ee
where we have used that in the zero-width limit ($\Gamma^{\rm tot}_\rho\to 0$), the expression in parentheses reduces
to~$\delta(s-m_\rho^2)$. Thus, we recover the LCSR for the $B\to \rho$ form factor $V^{B\rho}(q^2)$ obtained
in~Ref.~\cite{Khodjamirian:2006st}. Analogously, starting from Eqs.~(\ref{eq:LCSRs-AV-g}), (\ref{eq:LCSRs-AV-qp})
and (\ref{eq:lcsrFtrans}) we recover the LCSRs for the $B\to\rho$ form factors $A_1^{B\rho}(q^2),A_2^{B\rho}(q^2)$
and $A_0^{B\rho}(q^2)$ respectively.

We note at this point that relating the form factor $A_3$ with the
form factors $A_{1,2}$ according to the relation quoted after \Eq{eq:BrhoFFdef}, and using the kinematic relation $A_0(0)=A_3(0)$, we obtain
an alternative sum rule for $A_0(0)$. This sum rule coincides with
\Eq{eq:rhodomin} up to $\op(\bar{\Lambda}/m_B)$ power corrections,
with $\bar{\Lambda}\sim m_B-m_b$, i.e., within the
usual accuracy of the LCSRs with $B$-meson DAs~\cite{Khodjamirian:2006st}.
Thus our sum rule for $A_0$ satisfies the kinematic relation $A_0(0) = A_3(0)$
up to power corrections.

\bigskip

Returning to the LCSRs in Eqs.~(\ref{eq:lcsrFperp})-(\ref{eq:LCSRs-AV-qp}) and~(\ref{eq:lcsrFtrans}) with a general hadronic representation of the dipion state, we note that these sum rules offer the opportunity to go beyond the $\rho$-dominance approximation and  to investigate the role of excited $\rho$ resonances in $B\to \pi\pi$ form factors. From the measurements of  the pion e.m. form factor in $e^+e^-$ annihilation (see e.g. Ref~\cite{Aubert:2009AD}) and the pion vector form factor in $\tau\to \pi^-\pi^0\nu_\tau$\cite{Fujikawa:2008ma},  it is known that in the region $s \lesssim 1.5~\GeV^2$  both form factors are accurately described by including, apart from the $\rho\equiv \rho(770)$, its two radial excitations: $\rho'\equiv\rho(1450)$ and  $\rho''\equiv\rho(1750)$~\cite{PDG}. In what follows, we adopt the three-resonance parametrization of $F_\pi(s)$  used by the Belle collaboration \cite{Fujikawa:2008ma} to fit their so far most accurate data on $\tau\to \pi^-\pi^0\nu_\tau$  (see Appendix \ref{app:pi-ff}). Various other parametrizations for the pion form factors in the timelike region can be found in the literature (see e.g. \cite{Bruch:2005PY,Czyz:2010hj,Achasov:2011RA,Hanhart:2012WI,Shekhovtsova:2012RA}). The important  point is that, at least in the low dipion-mass region, the ``nonresonant'' contributions can be described well by the interference of the $\rho$ with the $\rho'$ and $\rho''$, and these excited states contribute at the level of $15\text{-}20\%$ to the total form factor~\footnote{
Note that at larger $s$ the infinite tail of vector resonances may influence the form factor
and it presumably has to be taken into account \cite{Dominguez:2001zu,Bruch:2005PY,Czyz:2010hj}.
}.

Assuming that the formation and hierarchy of $\rho$-resonances in the $B\to \pi\pi$ transition is similar to that in $F_\pi(s)$, we adopt a three-resonance ansatz generalizing \Eq{eq:VrelFperp} for all vector and axial-vector $P$-wave form factors.
For the form factors $F^{(\ell=1)}_\bot $and $F^{(\ell=1)}_\|$ we write\,\footnote{
For the sake of generality, we include also the relative phase of the $\rho$ term.
}:
\bea
\label{eq:Bpipiansatz}
F_{\perp}^{(\ell=1)}(s,q^2)&=&\frac{\sqrt{s}
\sqrt{\lambda}}{\sqrt{3}}
\sum_{R}\frac{g_{R\pi\pi}V^{BR}(q^2)\ e^{i\phi_R{(s,q^2)}}}
{(m_B+m_R)\big[m_R^2-s-i\sqrt{s}\ \Gamma_R(s)\big]}\,,\\
F_{\parallel}^{(\ell=1)}(s,q^2)&=&
\frac{\sqrt{s}}{\sqrt{3}}
\sum_{R}\frac{(m_B+m_R) g_{R\pi\pi}A_1^{BR}(q^2)\ e^{i\phi_R{(s,q^2)}}}
{\big[m_R^2-s-i\sqrt{s}\ \Gamma_R(s)\big]}\,,
\label{eq:Bpipiansatz-1}
\eea
where the sum runs over $R = \{\rho,\rho',\rho''\}$.
The linear combination of $F^{(\ell=1)}_\|$ and $F^{(\ell=1)}_0$  entering the LCSR in \Eq{eq:LCSRs-AV-qp} is related
to the  $B\to R$ form factors $A_2^{BR}$:
\be
\frac{(m_B^2-s-q^2)}{\sqrt{s}}F_{\parallel}^{(1)}(s,q^2)
+\frac{2\sqrt{ q^2}}{\beta_{\pi}(s) }
F_{0}^{(1)}(s,q^2)
= \frac{\lambda}{\sqrt{3}}\sum_{R}\frac{g_{R\pi\pi}A_2^{BR}(q^2)\ e^{i\phi_R{(s,q^2)}}}
{(m_B+m_R)\big[m_R^2-s-i\sqrt{s}\ \Gamma_R(s)\big]}\,.
\label{eq:Bpipiansatz-2}
\ee
Finally, for the form factor $F_{t}^{(\ell=1)}$, we have
\bea
F_{t}^{(\ell=1)}(s,q^2)&=&
-\frac{\beta_{\pi} (s)\sqrt{\lambda}}{\sqrt{3}\sqrt{q^2}}\sum_{R}\frac{m_{R} g_{R\pi\pi}
A_0^{BR}(q^2)\ e^{i\phi_R{(s,q^2)}}}{\big[m_R^2-s-i\sqrt{s}\ \Gamma_R(s)\big]}\,.
\label{eq:Bpipiansatz-3}
\eea
For our exploratory study we refrain from using the  more involved resonance representation of \Ref{Gounaris:1968mw},
adopting instead a simpler Breit-Wigner  approximation with an energy-dependent width~\cite{Kuhn:1990ad}.
We also tacitly assume that the phase factors $\phi_R$ of resonance 
contributions are  independent of the form factor type.
On the other hand, it is conceivable that $q^2$-dependence 
of the $B\to R$ form factors is different for $R=\rho,\rho',\rho''$. 
Hence the simplest way to enforce the imaginary part condition of Eq.~(\ref{eq:phaseeq})
is to assume that this condition holds separately for each resonance term in the models
of Eqs.~(\ref{eq:Bpipiansatz})-(\ref{eq:Bpipiansatz-3}).
Then the phase $\phi_R$ is $s$-dependent but $q^2$-independent, and   
the general condition~(\ref{eq:phaseeq}) is replaced with the following relation specific to our resonance model:
\eq{
\tan[\delta_\pi(s)-\phi_R(s)]= \frac{\sqrt{s}\ \Gamma_R(s)}{m_R^2-s}\ ,
\qquad
\text{with}
\quad
F_\pi(s)=|F_\pi(s)|e^{i\delta_\pi(s)}\ .
\label{eq:phasecond2}
}
This relation essentially restricts the resulting phase dependence
of the $B\to \pi\pi$  form factors, in full analogy
with the well-known situation for  the timelike pion form factor.
Adopting a more general condition than \Eq{eq:phasecond2},
would enforce an artifitial compensation of $q^2$-dependences
in the phases $\phi_R(s,q^2)$ and $B\to R$ form factors,
in order to formally obey \Eq{eq:phaseeq}.
Moreover, there are several  physical arguments
in favour of $q^2$-independent phases of $B\to \pi\pi$ form factors
in the  region of dipions with large recoil and small invariant mass.
First, varying $q^2$ corresponds to varying
the total energy of the dipion state,
produced in the $B$-meson rest frame, whereas
the (Lorentz-invariant) amplitude of the final-state strong
interaction developing the phase depends only on the invariant
mass $s$ of the dipion. Second, similar to the factorization
in nonleptonic $B$-decays to light hadrons, the
hadronization and related strong interaction of 
the fast dipion system in the $B$-meson rest frame takes place
beyond the weak $b\to u$ transition domain. 

We parametrize the $q^2$-dependence of the $B\to R$ form factors entering Eqs.~(\ref{eq:Bpipiansatz})-(\ref{eq:Bpipiansatz-3}) with the standard $z$-series expansion~\cite{BCL}, in the form adopted in \Ref{KMPW}. 
The $z$-parametrization of the momentum transfer is given by
\be
z_R(q^2)=\frac{\sqrt{t_+^{R}-q^2}-\sqrt{t_+^{R}-t_0^{R}}}{\sqrt{t_+^{R}-q^2}+\sqrt{t_+^{R}-t_0^{R}}}\,,
\label{eq:zdef}
\ee
where $t_{\pm}^{R} \equiv (m_B \pm m_R)^2$ and $t_0^{R}=t_+^{R}(1-\sqrt{1-t_-^{R}/t_+^{R}})$.
For a generic form factor ${\cal F}^{BR}(q^2)$,
where ${\cal F} = \{V,A_1,A_2,A_0\}$ and $R=\rho,\rho',\rho''$, 
we have:
\be
{\cal F}^{BR}(q^2)= \frac{{\cal F}^{BR}(0)}{1-q^2/m^2_{{\cal F}}}
\Big\{ 1 + b^R_{\cal F}\,\zeta_R(q^2) + \cdots \Big\}\,,
\label{eq:zpar}
\ee
where we use the shorthand notation
\begin{equation}
\zeta_R(q^2)=z_R(q^2) - z_R(0) + \frac12 [z_R(q^2)^2-z_R(0)^2] \,,
\nonumber
\end{equation}
and 
$m_{{\cal F}}$ is the lowest heavy-light pole mass in the $q^2$ channel
with spin-parity depending on the type of the form factor.

We now substitute the resonance models of  Eqs.~(\ref{eq:Bpipiansatz})-(\ref{eq:Bpipiansatz-3})
into the sum rules (\ref{eq:lcsrFperp})-(\ref{eq:LCSRs-AV-qp}) and~(\ref{eq:lcsrFtrans}) respectively.
For the sake of brevity we introduce the following notation:
\bea
&&\kappa_{\cal F}^R\equiv g_{R\pi\pi}{\cal F}^{BR}(0)\ , \quad
\eta_{\cal F}^R \equiv g_{R\pi\pi}{\cal F}^{BR}(0)b_{\cal F}^R\ ,
\nonumber\\[1mm]
&& X_V^R = X_{A_2}^R = (m_B+m_R)^{-1}\ ,\quad X_{A_1}^R = (m_B+m_R)\ , \quad X_{A_0}^R = -m_R\ ,
\label{eq:parfit}
\eea
so that all four  sum rules can be rewritten in a generic form:
\eq{
\sum_R
\frac{\kappa_{\cal F}^R+\eta_{\cal F}^R \ \zeta_R(q^2)}{1-q^2/m^2_{{\cal F}}}\ 
X_{\cal F}^R\ I_R(s_0^{ 2\pi},M^2)
\,  
= I^{\rm OPE}_{\cal F}(s_0^{{ 2\pi}},M^2,q^2)\,.
\label{eq:fitrel}
}
In the above, the functions $I^{\rm OPE}_{\cal F}(s_0^{{ 2\pi}},M^2,q^2)$
with ${\cal F} = \{V,A_1,A_2,A_0\}$
represent the r.h.s of Eqs.~(\ref{eq:lcsrFperp})-(\ref{eq:LCSRs-AV-qp}) and (\ref{eq:lcsrFtrans}) respectively
and the coefficients of the $B\to R$ form factors 
are given by the integrals:
\eq{
I_R(s_0^{ 2\pi},M^2) =  \frac{1}{12\sqrt{2}\pi^2}
\int\limits_{4m_{\pi}^2}^{s_0^{2\pi}} ds 
\,e^{-s/M^2}\frac{ s \ [\beta_{\pi}(s)]^3 \ |F_{\pi}(s)|}{\sqrt{(m_R^2-s)^2+s\,\Gamma_R^2(s)}} 
\,.
\label{eq:sInt}
}
The set of sum rule relations (\ref{eq:fitrel}) can be used 
to fit the parameters $\kappa_{\cal F}^R$ and  $\eta_{\cal F}^R$  of the resonance models 
in Eqs.~(\ref{eq:Bpipiansatz})-(\ref{eq:Bpipiansatz-3}) for the $B\to \pi\pi$ form factors.

\section{Input and numerical analysis}\label{numerics}

The input in the LCSRs~(\ref{eq:lcsrFperp})-(\ref{eq:LCSRs-AV-qp}) and (\ref{eq:lcsrFtrans}) includes the parameters of $B$-meson DAs described in Appendix~\ref{app:B-DA}.
The most important one is the inverse moment $\lambda_B$,
which has still a rather large uncertainty. The interval
\begin{equation}
\lambda_B\equiv \lambda_B(1 \GeV)= 460\pm 110 ~\mbox{MeV}\,,
\label{eq:lambdaB}
\end{equation}
predicted from QCD sum rules~\cite{Braun:2003wx}, is
in agreement with the  lower limit $\lambda_B> 238$~MeV (at $90\%$~C.L.) recently obtained
by the Belle collaboration~\cite{Heller:2015vvm}, combining the search for $B\to \gamma\ell \nu_\ell$ with the theory prediction for its branching ratio~\cite{Beneke:2011nf,Braun:2012kp}. This limit is starting to challenge
the lower values around $\lambda_B=200-250 $~MeV preferred by the QCD factorization analysis
of $B\to \pi\pi$ nonleptonic decays  (see e.g., Refs.~\cite{Beneke:2015wfa,Beneke:2009ek}).
In addition, a recent estimate $\lambda_B= 358^{+38}_{-30}$~MeV~\cite{Wang:2015VGV} has been obtained by comparing the LCSRs with pion~\cite{Khodjamirian:2011ub} and $B$-meson DAs for the $B\to\pi$ form factor
and using the same model for the $B$-meson DA as the one used here.
In our numerical analysis we adopt the central value and uncertainty
of the sum rule prediction quoted in Eq.~(\ref{eq:lambdaB}).

Since we do not include NLO corrections in the correlation functions,
we also do not take into account the renormalization of $\lambda_B$.
In the absence of perturbative corrections, 
the choice of renormalization scale for the correlation function 
remains an open issue. This choice concerns 
especially the sum rule (\ref{eq:lcsrFtrans}) for which 
the $b$-quark mass  is needed.
We choose a typical  $\overline{\rm MS}$ value $m_b(m_b)=4.2~\GeV $.
The value of the  (scale-independent)  $B$-meson decay constant entering the OPE part 
of all LCSRs is known with a reasonable accuracy from the 2-point QCD sum rules. We use
 $ f_B= 207^{+17}_{-9}$~MeV from Ref.~\cite{Gelhausen:2013wia}, which agrees well 
with lattice QCD determinations~\cite{FLAG}.

For the Borel parameter $M^2$ in all the sum rules we take values inside the interval \mbox{$M^2=1.0 - 1.5~\GeV^2$},
which is slightly narrower  than the one used in \Ref{Khodjamirian:2005ea}.
For this interval, the convergence of OPE is manifested by relatively small three-particle  DA contributions
(with $I_{\cal F}^\text{OPE, 3-particle}/I_{\cal F}^\text{OPE, 2-particle}\lesssim 20\%$ at $q^2=0$
and $\lesssim 30\%$ at $q^2=10\GeV^2$).
Simultaneously, the integral over the spectral density of the correlation function (r.h.s. of \Eq{eq:dual})
does not exceed 40\% of the total integral, making the result not too sensitive to the quark-hadron duality approximation.

The choice of the threshold parameter $s_0^{2\pi}$ deserves a separate investigation.
We emphasize that here a quark-hadron duality pattern is used that is more involved  than the
conventional one-pole-plus-continuum ansatz.
Hence, for consistency we fix the threshold employing the two-point SVZ sum rule~\cite{Shifman:1979BX} for the
isospin-one light-quark vector currents, where we substitute the pion timelike form factor in the hadronic part.
The details are given in Appendix~\ref{app:2pt-s0}.
The result depends mildly on the value of the Borel parameter $M^2$, and within our
chosen range it leads to $s_0 \simeq 1.5 \GeV^2$ quite generically, in the same ballpark as the one obtained with the
one-pole ansatz.

The remaining input concerns the hadronic parameters in Eq.~(\ref{eq:sInt}).
For the pion form factor $F_\pi(s)$ we use the model of \Ref{Fujikawa:2008ma} and the fit results for its parameters given in that paper, which are collected in Appendix~\ref{app:pi-ff} for convenience. These include determinations for $m_R$ and $\Gamma_R$ appearing explicitly in Eq.~(\ref{eq:sInt}).
For the pole masses $m_{\cal F}$ in Eqs.~(\ref{eq:zpar}) and~(\ref{eq:fitrel}),
we use~\cite{PDG}:
\eqa{
m_V=m_{B^*} &=&5.325~\GeV\ \ (J^P=1^-)\ ,
\nonumber\\
m_{A_{1,2}}=m_{B_1}&=&5.726~\GeV\ \ (J^P=1^+)\ ,
\\
m_{A_0}=m_B&=&5.279 \GeV\ \ (J^P=0^-)\ .\nonumber
}

After specifying all input parameters, we calculate the coefficients 
$I_{R}$ and $I_{\cal F}^\text{OPE}$  in \Eq{eq:fitrel}.
In order to have an impression on the relative contributions of all three
resonances to the sum rule relations (\ref{eq:fitrel})  we quote the values of the coefficients $I_R$,
varying the values for all masses, widths, and the parameters in $F_\pi$ as given in Table~\ref{TableFpiBelle}: 
\eq{
I_\rho = (26\pm 3) \cdot 10^{-3}\ ,\quad
I_\rho' = ( 4.3 \pm 0.8) \cdot 10^{-3}\ ,\quad
I_\rho'' = ( 2.5 \pm 0.5) \cdot 10^{-3}\ ,\quad
\label{eq:Ival}
}
calculated at  $s_0^{2\pi}=1.5$ GeV$^2$ and $M^2=1.0$ GeV$^2$.
We also quote here the value of the strong coupling $g_{\rho\pi\pi}$ derived from Eq.~(\ref{eq:width}):
\eq{
g_{\rho\pi\pi} = 5.96\pm 0.04\ ,
\label{eq:gRpipi}
}
which is necessary to relate the coefficients $\kappa_{\cal F}^\rho$ to the form factors ${\cal F}^{B\rho}$.
In the following we will consider all uncertainties entering $I_{\cal F}^\text{OPE}$, but fix the
hadronic parameters entering $I_{R}$ and $g_{\rho\pi\pi}$ to their central values.

\subsection{Finite-width effects in $B\to\rho$ form factors}

We start our numerical analysis reproducing the results of \Ref{Khodjamirian:2006st} for the $B\to \rho$ form factors
by taking the one-pole ansatz for the $B\to \pi\pi$ form factors,
i.e. retaining only the $\rho$ in Eqs.~(\ref{eq:Bpipiansatz})-(\ref{eq:Bpipiansatz-3})
in the narrow width approximation,
and subsequently taking into account the corrections arising from the finite
width of the $\rho$ and the effect from higher resonances (acting effectively as a
``nonresonant" background). We do this in several steps, with the results summarized in Table~\ref{tablerho}:

\bigskip

Employing the one-resonance models in Eqs.~(\ref{eq:VrelFperp}),~(\ref{eq:VrelFpi})
-- and the analogous models for $F_{\|,0,t}$-- in the limit $\Gamma_\rho^\text{tot}\to 0$,
we use the sum rules to calculate the form factors $V^{B\rho}(q^2)$ and $A^{B\rho}_{1,2,0}(q^2)$ at $q^2=0$.
With the same inputs as used in \Ref{Khodjamirian:2006st} we find
good agreement with the central values quoted in that paper\footnote{
The form factor $A_0^{B\rho}$ was not calculated in \Ref{Khodjamirian:2006st}.
Thus the results for $A_0^{B\rho}$ given here are new.
}. These numbers are collected in the first row of
Table~\ref{tablerho}. Updating the input parameters to
the ones quoted at the beginning of this section (but still keeping $M^2=1.0 \GeV^2$ fixed),
we find a $\sim 10\%$ enhancement in the central values, due mostly to the change in the numerical input
for the $B$-meson decay constant: $f_B = 180\MeV \to 207 \MeV$ (second row of Table~\ref{tablerho}).
Performing a gaussian scan over the parameters with uncertainties, we calculate central values and
errors for the form factors by taking the mean and standard deviation of the resulting distributions for the
form factors (third row of Table~\ref{tablerho}). This shifts the central values further up slightly (but well within
the uncertainties). The larger error bars with respect to \Ref{Khodjamirian:2006st} are due to the different
approach used here (gaussian versus flat scans). These are our results in the single-pole approximation
($\rho$-dominance, zero-width). For reference we show in the fourth row in Table~\ref{tablerho} the results
for the form factors obtained in \Ref{BSZ} (updating \Ref{Ball:2005RG}) using the LCSRs with $\rho$-meson DAs,
in which the zero-width approximation is also adopted. 

\bigskip

We now maintain the one-resonance model for the $B\to\pi\pi$ form factors, but
adopt the full Belle~\cite{Fujikawa:2008ma} data-based model for $F_\pi$ (see Appendix~\ref{app:pi-ff}).
Keeping $M^2=1.0\GeV^2$, we obtain the results quoted in the fourth row of Table~\ref{tablerho}, which imply a
$\sim 10\%$ enhancement with respect to the zero-width limit. Our final results for the single resonance
model are obtained by simultaneously fitting the sum rules with different values of the Borel parameter
$M^2 =\{1.0,1.25,1.5\} \GeV^2$. The results are given in the last row of Table~\ref{tablerho}. The central values
are essentially unchanged,
but the uncertainties are reduced because each value of $M^2$ acts as a separate constraint.

\bigskip

We conclude that the finite width of the $\rho$ and the presence of higher resonances in $F_\pi$
impact the $B\to\rho$ form factors at the level of $\sim 10\ {\rm to}\ 15\%$, when
the $B\to\rho$ form factors are defined from the $B\to\pi\pi$ form factors by neglecting the contributions from excited resonances in Eqs.~(\ref{eq:Bpipiansatz})-(\ref{eq:Bpipiansatz-3}). 
This is in agreement with the findings in \Ref{Hambrock:2015aor}.

\begin{table}
\centering
\begin{tabular}{@{}rccccc@{}}
\toprule
&\hspace{0mm} & $V^{B\rho}(0)$ & $A_1^{B\rho}(0)$ & $A_2^{B\rho}(0)$ & $A_0^{B\rho}(0)$ \\
\midrule
Inputs of \Ref{Khodjamirian:2006st} && $0.31$ & $0.23$ & $0.19$ & $0.26$ \\
Updated inputs && $0.34$ & $0.26$ & $0.21$ & $0.30$ \\
Gaussian scan && $0.36 \pm 0.17$ & $0.27 \pm 0.13$ & $0.22 \pm 0.15$ & $0.30 \pm 0.06$\\[1mm]
\hdashline[1pt/2.3pt]\\[-6mm]
\Ref{BSZ} ($\rho$-DAs)&& $0.33 \pm 0.03$ &  $0.26 \pm 0.03$ & $0.23 \pm 0.04$ &  $0.36 \pm 0.04$\\
\midrule
Full $F_\pi$, $M^2=1\GeV^2$ && $0.40\pm 0.19$ & $0.30\pm 0.14$ & $0.24\pm 0.16$ & $0.33\pm 0.07$\\
Final results for $\rho$-model && $0.41 \pm 0.11$ & $0.31 \pm 0.08$ & $0.25 \pm 0.10$ & $0.34 \pm 0.04$\\
\bottomrule
\end{tabular}
\caption{\it Results for the $B\to\rho$ form factors in the one $\rho$-resonance model for the $B\to\pi\pi$ form factors.
The first four rows correspond to the zero-width approximation, while the last two rows
include finite-width effects.}
\label{tablerho}
\end{table}

\subsection{Assessing the $\rho'$ contribution to $B\to\pi\pi$ form factors}

In order to estimate the $\rho'$ contribution, we now assume that the $B\to\rho$
form factors are well determined from the LCSRs with 
$\rho$-meson DAs, obtained in the narrow-$\rho$ approximation.
We thus take the models in Eqs.~(\ref{eq:Bpipiansatz})-(\ref{eq:Bpipiansatz-3}), neglecting the $\rho''$ contribution, and use the
results from \Ref{BSZ} to fix the parameters
$\kappa_{\cal F}^\rho$ and $\eta_{\cal F}^\rho$. The free parameters in the resulting models are then only
$\kappa_{\cal F}^{\rho'}$ and $\eta_{\cal F}^{\rho'}$.

We then use the sum rules~(\ref{eq:lcsrFperp})-(\ref{eq:LCSRs-AV-qp}) and (\ref{eq:lcsrFtrans}) to determine these
parameters. Besides using, as in the previous section, three different
values for the Borel parameter $M^2 =\{1.0,1.25,1.5\} \GeV^2$, we consider various $q^2$ points:
$q^2 = \{ 0, 1, \dots, 10\} \GeV^2$, in order to determine the slope parameters~$\eta_{\cal F}^{\rho'}$.
The results of this fit are shown in Table~\ref{tablerhoprime}. Due to the suppressed sensitivity of the sum rules
to the $\rho'$ region (see \Eq{eq:Ival}), the uncertainties on the parameters $\kappa_{\cal F}^{\rho'}$ and
$\eta_{\cal F}^{\rho'}$ are rather large. Thus our fit allows for a quite appreciable $\rho'$ contribution relative
to the (fixed) $\rho$ contribution.

\begin{table}
\centering
\begin{tabular}{c@{}cr@{}c@{}l c r@{}c@{}l c r@{}c@{}l c r@{}c@{}l c r@{}c@{}l@{}}
\toprule
&\hspace{5mm} && $V$ &&&& $A_1$ &&&& $A_2$ &&&& $A_0$ & \\
\midrule
$\kappa_{\cal F}^{\rho}$     &&   $2.0$ & $\,\pm\,$ & $0.2$       && $1.6$ & $\,\pm\,$ & $0.2$       && $1.4$ & $\,\pm\,$ & $0.2$         && $2.1$ & $\,\pm\,$ & $0.2$ \\
$\eta_{\cal F}^{\rho}$       &&   $-5.1$ & $\,\pm\,$ & $1.1$     && $2.3$ & $\,\pm\,$ & $0.8$       && $-2.8$ & $\,\pm\,$ & $1.2$         && $-5.0$ & $\,\pm\,$ & $1.2$ \\
\midrule
$\kappa_{\cal F}^{\rho'}$     &&   $3.0$ & $\,\pm\,$ & $2.5$       && $1.5$ & $\,\pm\,$ & $1.4$       && $1.0$ & $\,\pm\,$ & $2.2$         && $-0.3$ & $\,\pm\,$ & $0.4$ \\
$\eta_{\cal F}^{\rho'}$       &&   $-52$ & $\,\pm\,$ & $74$     && $-2$ & $\,\pm\,$ & $35$       && $26$ & $\,\pm\,$ & $65$         && $-8$ & $\,\pm\,$ & $12$ \\
correlation                &&                   & + & $0.8$     &&               & + & $0.9$       &&               & + & $0.8$         &&                & + & $0.8$ \\
\bottomrule
\end{tabular}
\caption{\it Results of the fit to the $\rho'$ contribution.}
\label{tablerhoprime}
\end{table}

\subsection{Three-resonance model}

We now consider the full three-resonance models given in Eqs.~(\ref{eq:Bpipiansatz})-(\ref{eq:Bpipiansatz-3}).
This model however contains too many parameters to be
independently fitted from the sum rules, in which the
contributions of $\rho',\rho''$ enter with suppressed coefficients
with respect to the $\rho$ contribution (see Eq.~(\ref{eq:Ival})). In the future, when
sufficient amount of data on $B\to \pi\pi \ell \nu_\ell$ is accumulated,
one should be able to isolate the $P$-wave dipions in this decay
and perform  a more refined analysis combining these data with the sum rule  constraints.
For the time being, the only information on the role of
higher resonances we have is provided by the $F_\pi$ measurement
given by the parametrization in Eq.(55). Note that in the pion
vector form factor the  $R$-resonance ($R=\rho,\rho',\rho'',\dots$) contribution is
determined by the product of the decay constant of $R$ and the
strong coupling $g_{R\pi\pi}$ , whereas in the $B\to \pi\pi$ form factor
the $R$-contribution to the resonance model is determined  by the
$B\to R$ form factor multiplied by the coupling $g_{R\pi\pi}$.
Owing to quite different physical processes, the ratios of $B\to R$ form factors may deviate considerably
from the ratios of $R$ decay constants.
E.g., it is plausible that at large recoil the $B\to \rho'$  transition
is even enhanced with respect to $B\to \rho$ transition
because the hadronization into a larger mass is more probable.
Nevertheless, since in this paper we want to illustrate numerically
the influence of the nonresonant  background in the $B\to \pi\pi$
transitions at small dipion mass, it is conceivable to assume
that the relative size of the contributions from $\rho,\rho'$ and $\rho''$
is the same as in $F_\pi$.
We do that by imposing the following conditions:
\eqa{
\kappa_{\cal F}^{\rho'} = \beta \,\kappa_{\cal F}^\rho\ ,
& \qquad &
\eta_{\cal F}^{\rho'} = \beta \, \left[ \lim_{q^2\to 0} \frac{\zeta_\rho(q^2)}{\zeta_{\rho'}(q^2)} \right] \,\eta_{\cal F}^\rho\ ,
\nonumber\\[2mm]
\kappa_{\cal F}^{\rho''} = \gamma \,\kappa_{\cal F}^\rho\ ,
& \qquad &
\eta_{\cal F}^{\rho''} = \gamma \, \left[ \lim_{q^2\to 0} \frac{\zeta_\rho(q^2)}{\zeta_{\rho''}(q^2)} \right] \,\eta_{\cal F}^\rho\ ,
\label{conditions3R}
}
where $\beta$ and $\gamma$ are the parameters in the parametrization of $F_\pi$ in Eq.~(\ref{FpiBelle}).
The conditions on $\kappa_{\cal F}^R$ fix the relative contributions at $q^2=0$ and the conditions on
$\eta_{\cal F}^R$ fix the derivatives. At the end we find that the conditions~(\ref{conditions3R}) fix the relative
contributions in the full $q^2$ region with good accuracy.

\begin{table}
\centering
\begin{tabular}{c@{}cr@{}c@{}l c r@{}c@{}l c r@{}c@{}l c r@{}c@{}l c r@{}c@{}l@{}}
\toprule
&\hspace{5mm} && $V$ &&&& $A_1$ &&&& $A_2$ &&&& $A_0$ & \\
\midrule
$\kappa_{\cal F}^\rho$     &&   $2.4$ & $\,\pm\,$ & $0.4$       && $1.8$ & $\,\pm\,$ & $0.3$       && $1.5$ & $\,\pm\,$ & $0.3$         && $1.9$ & $\,\pm\,$ & $0.1$ \\
$\eta_{\cal F}^\rho$       &&   $-11.0$ & $\,\pm\,$ & $8.4$     && $2.0$ & $\,\pm\,$ & $4.9$       && $0.2$ & $\,\pm\,$ & $7.3$         && $-6.5$ & $\,\pm\,$ & $2.8$ \\
correlation                &&                   & + & $0.8$     &&               & + & $0.9$       &&               & + & $0.8$         &&                & + & $0.8$ \\[1mm]
${\cal F}^{B\rho}(0)$      &&   $0.39$ & $\,\pm\,$ & $0.06$     && $0.30$ & $\,\pm\,$ & $0.04$     && $0.25$ & $\,\pm\,$ & $0.05$       && $0.32$ & $\,\pm\,$ & $0.02$\\
\midrule
$\kappa_{\cal F}^{\rho'}$  &&   $0.35$ & $\,\pm\,$ & $0.06$     && $0.27$ & $\,\pm\,$ & $0.04$      && $0.22$ & $\,\pm\,$ & $0.05$      && $0.29$ & $\,\pm\,$ & $0.02$ \\
$\eta_{\cal F}^{\rho'}$    &&   $-2.11$ & $\,\pm\,$ & $1.62$    && $0.38$ & $\,\pm\,$ & $0.94$      && $0.03$ & $\,\pm\,$ & $1.41$      && $-1.25$ & $\,\pm\,$ & $0.54$ \\
\midrule
$\kappa_{\cal F}^{\rho''}$ &&   $0.09$ & $\,\pm\,$ & $0.01$     && $0.07$ & $\,\pm\,$ & $0.01$      && $0.05$ & $\,\pm\,$ & $0.01$      && $0.07$ & $\,\pm\,$ & $0.00$ \\
$\eta_{\cal F}^{\rho''}$   &&   $-0.57$ & $\,\pm\,$ & $0.44$    && $0.10$ & $\,\pm\,$ & $0.25$      && $0.01$ & $\,\pm\,$ & $0.38$      && $-0.34$ & $\,\pm\,$ & $0.14$ \\
\bottomrule
\end{tabular}
\caption{\it Results of the three-resonance fit.}
\label{table3R}
\end{table}

These simplified models depend only on two parameters for each form factor:
$\kappa_{\cal F}^\rho$ and $\eta_{\cal F}^\rho$. We use the sum rules to determine these parameters, using again
the three values of $M^2 =\{1.0,1.25,1.5\} \GeV^2$, and also $q^2 = \{ 0, 1, \dots, 10\} \GeV^2$. The results of
the fit are given in Table~\ref{table3R}. We note that the values for $\kappa_{\cal F}^\rho$ and $\eta_{\cal F}^\rho$
are strongly correlated within each form factor, with correlation coefficients given in the third row.
We provide for completeness also the resulting form factors $V^{B\rho},A_{1,2,0}^{B\rho}$
at $q^2=0$ and the values of the parameters $\kappa_{\cal F}^R$ and $\eta_{\cal F}^R$ for $R=\rho',\rho''$,
although all these numbers can be obtained rather trivially from the first three rows.

Given the fact that the $\rho'$ and $\rho''$ contributions to the pion form factor are relatively small,
the results of the $B\to\rho$ form factors obtained in the constrained three-resonance model are
in good agreement with the ones obtained in the $\rho$-model (compare the third row in Table~\ref{table3R} with
the fifth row in Table~\ref{tablerho}). The effect of the higher $\rho^{\prime,\,\prime\prime}$ resonances is to
decrease slightly the $B\to\rho$ form factors. The uncertainties obtained in this section are smaller only because
the fit includes many points in the full $q^2$ region, all acting as separate (and consistent) constraints,
while in Table~\ref{tablerho} we only considered the form factors at $q^2=0$.

\begin{figure}[t]
\includegraphics[width=\textwidth]{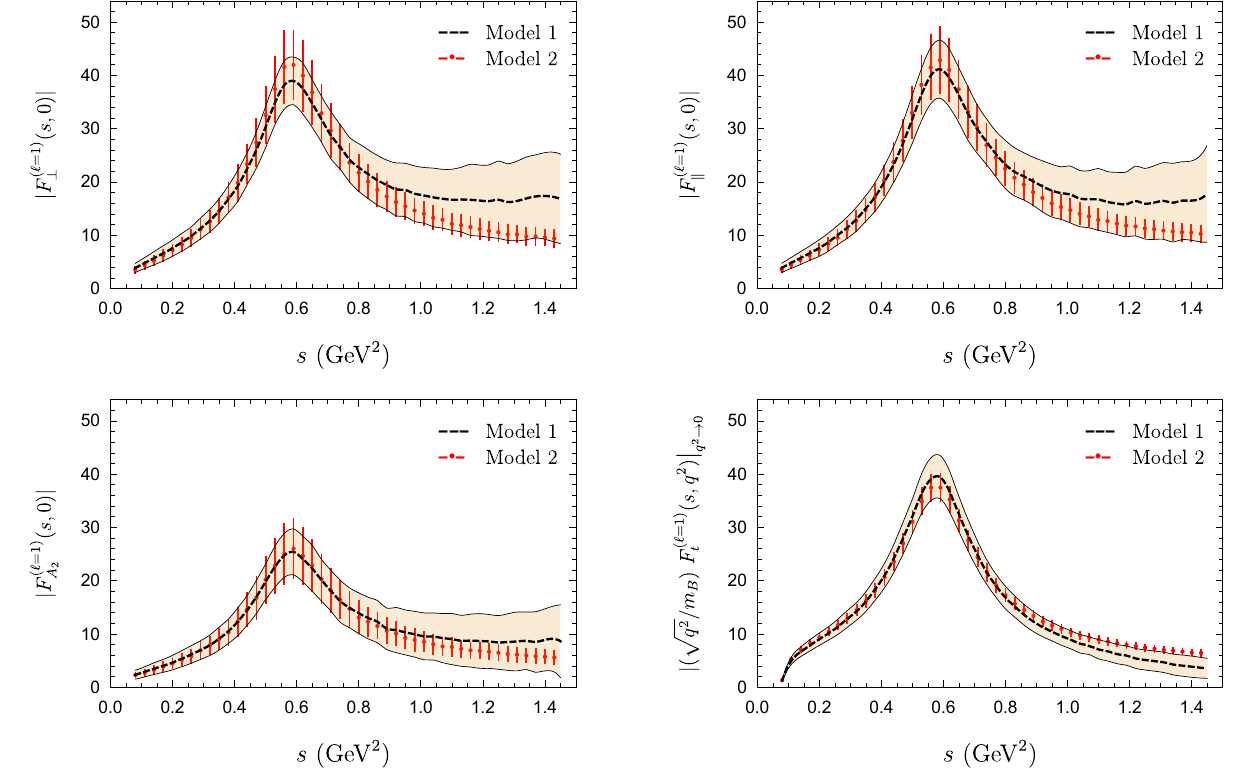}
\caption{\it $B\to \pi\pi$ form factors at $q^2=0$ as a function of the dipion mass,
within the models of Table~\ref{tablerhoprime} (Model 1) and Table~\ref{table3R} (Model 2).
Shaded regions and error bars account for all uncertainties.}
\label{figplots}
\end{figure}

In Fig~\ref{figplots} we show the results for the absolute values of the form factors,
comparing the model results in Table~\ref{tablerhoprime} (Model 1) and Table~\ref{table3R} (Model 2).
The results for $F_0^{(\ell =1)}$ depend on the correlations between $\kappa^R_{A_1}$ and $\kappa^R_{A_2}$, and in order to ignore this correlation we plot instead $F_{A_2}^{(\ell =1)}$ defined as
\eq{
F_{A_2}^{(\ell =1)}(s,q^2) \equiv
\frac{m_B^2 - s - q^2}{m_B^2} F_\|^{(\ell =1)}(s,q^2)
+ \frac{2 \sqrt{s q^2}}{\beta_\pi(s) m_B^2} F_0^{(\ell = 1)}(s,q^2)\ ,
}
which at $q^2=0$ depends only on $\kappa^R_{A_2}$. There is good agreement between both models.
Due to large uncertainties in $\kappa^{\rho'}_{\cal F}$, Model~1 yields broader intervals for the form factors at $s$ above
the $\rho$ region. Larger form factors in this region are compensated by slightly smaller values around $s\sim m_\rho^2$
in order to satisfy the sum rules.

\begin{figure}[t]
\includegraphics[width=\textwidth]{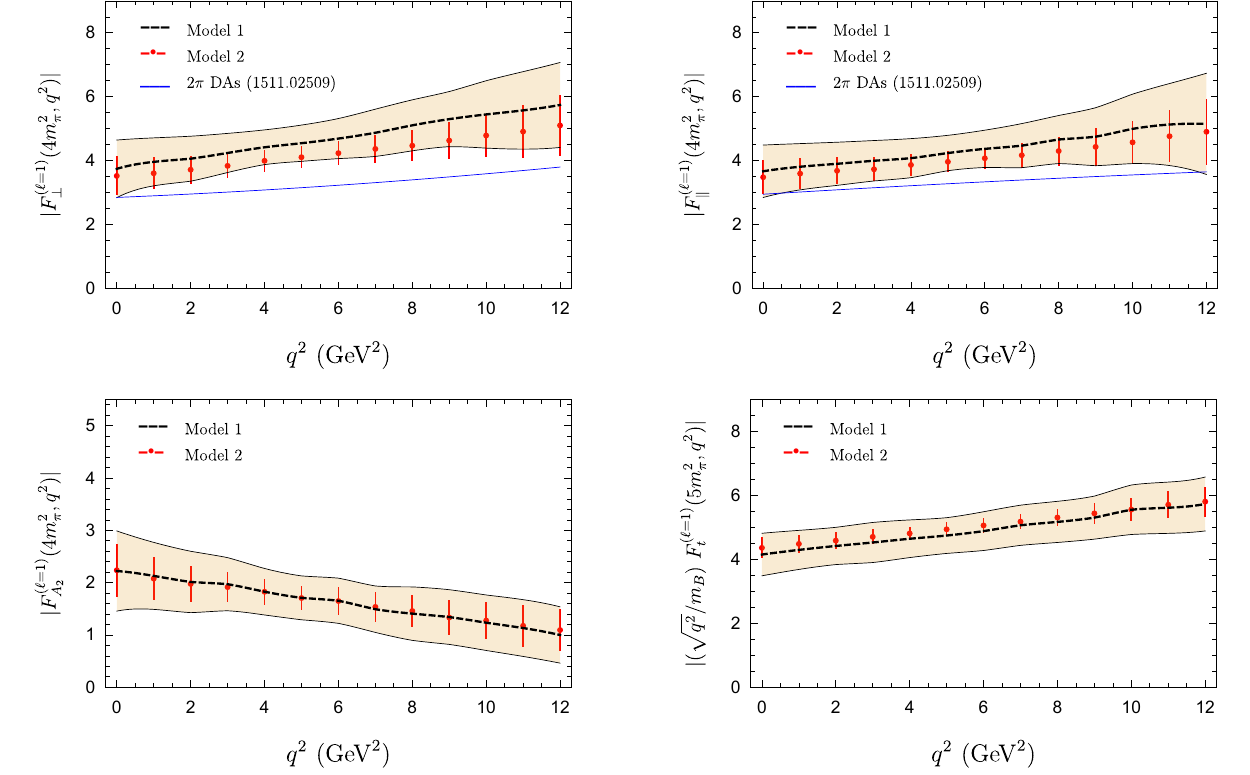}
\caption{\it $B\to \pi\pi$ form factors as a function of the momentum transfer $q^2$ at the dipion threshold $s=4m_\pi^2$
($F_t$ is plotted instead at $s=5m_\pi^2$).
The models of Table~\ref{tablerhoprime} (Model~1) and Table~\ref{table3R} (Model~2) are compared to the predictions
for $F_{\bot,\|}$ from the LCSRs with dipion DAs given in \Ref{Hambrock:2015aor} (only central values).
Shaded regions and error bars account for all uncertainties.}
\label{figplots2}
\end{figure}

Within Model 1 (Table~\ref{tablerhoprime}), the fitted intervals for $B\to \rho' $ form factors (albeit with very large uncertainties) do not exclude a noticeable (up to 20\%) contribution  of the $B\to \rho'$  transition to the  total budget of the $B\to \pi\pi$ form factors at small dipion masses.
In Model~2 (Table~\ref{table3R}) where the relatively small 
(most probably underestimated) contributions of both $\rho'$ and $\rho''$ are fixed, 
the resulting $B\to \rho$ form factor grows insignificantly 
with respect to the one in Model 1, staying within the estimated
uncertainties of the latter.  We conclude that the LCSRs with $B$ meson DAs
provide a stable prediction for the dominant $B\to \rho$  part 
of the $P$-wave $B\to \pi\pi$ form factors, provided the $B\to\rho'$ component
remains bounded. The transitions to 
excited $\rho$-mesons, being subdominant in the sum rules, 
cannot be predicted with a high degree of accuracy 
unless one adopts some particular ansatz for their pattern.
At the same time a sizeable contribution from excited states is consistent with
our LCSRs, and is supported by the independent LCSRs in terms of dipion DAs considered in \Ref{Hambrock:2015aor}.
Hence, in the future, more precise 
measurements of $B\to \pi\pi$  form factors must include these
contributions with interfering phases in their fits. This is 
a necessary step in accurately determining the $B\to \rho$ form factors.
Restricting  the dipion mass in the $\rho$-mass region, as it is usually 
done (see e.g.~\Ref{Sibidanov:2013rkk}), is not sufficient 
if an accuracy better than 15-20\% is sought.

We finish this section with a brief discussion on the $q^2$ dependence of the form factors, and comparing our results
to the ones obtained in \Ref{Hambrock:2015aor}. This is shown in Fig.~\ref{figplots2}. We find that
our results are compatible with the results of \Ref{Hambrock:2015aor} for $F_{\bot,\|}$, with the absolute magnitude
of the latter a bit below our results. The calculation of $F_{0,t}$ in terms of dipion DAs is still work in progress
(only a relationship between $F_t$, $F_0$ and $F_\|$ is given in \Ref{Hambrock:2015aor}), and thus we cannot
perform such comparison for $F_{A_2}$ and $F_t$.
Our results for the slope parameters $\eta_{\cal F}$ have a significant error, and thus one would naively expect
that the uncertainties in the form factors increase visibly with $q^2$, contrary to what is seen in Fig.~\ref{figplots2}
where the uncertainties are rather constant. The reason for this is the large positive correlation between $\kappa_{\cal F}$
and $\eta_{\cal F}$ (of around $+0.8$, see Tables~\ref{tablerhoprime} and~\ref{table3R}). Since $\zeta_R(q^2)$ is
negative, lower values of $\kappa_{\cal F}$ (corresponding to lower values of the form factors at $q^2=0$)
are correlated with lower values of $\eta_{\cal F}$ (corresponding to larger slopes for the form factors and larger
values at $q^2=12 \GeV^ 2$), and viceversa.

\section{Conclusions}
\label{conclusion}

In this paper we have suggested a new approach to 
the $B \to \pi\pi$ transition form factors in the region of small dipion mass, 
employing the LCSRs  with $B$-meson DAs.
We have focused on the particular $\bar{B}^0\to \pi^+\pi^0$ transition,
generated by the weak $b\to u$ current,
with an isospin-one and $P$-wave dipion final state. The fact that this state
is interpolated by the light-quark vector current allows one to go beyond the single narrow $\rho$-meson approximation, probing  also the contributions of other intermediate
states with the same quantum numbers. We have obtained the LCSRs in a  general 
form in which the convolutions of $B\to 2\pi$ $P$-wave form factors 
with the pion vector form factor  
integrated over the quark-hadron duality interval are related to 
the integrals over $B$-meson DAs.
The latter are calculated from the OPE of the underlying correlation function,
taking into account the two- and three-particle $B$-meson DAs and reproducing the results
obtained earlier in  Refs.~\cite{Khodjamirian:2005ea,Khodjamirian:2006st}. In addition, we have derived a new
sum rule for the form factor $F_t$ starting from a slightly modified correlation function
and including the three-particle contribution $\Delta A_0^{B\rho}$, which is given here explicitly
for the first time (see Appendix~\ref{app:B-DA}).

We have performed an exploratory numerical analysis using as an input the vector form factor measured by the Belle collaboration,
and fitted in a form of a superposition of three resonances.
We have then investigated the impact of the nonresonant and excited states on the sum-rule results for the dominant $B\to \rho$ form factor,
including the effects of the total width of $\rho$, and of the excited resonance contributions to the vector form factor.
Using an independent calculation of $B\to \rho$ form factors from LCSRs with  zero-width  $\rho$-meson DAs,
we find that the contributions from $\rho'$ and other states in the region of low dipion mass can be typically at the
level of 15-20\%. This is consistent with the results of \Ref{Hambrock:2015aor} based on LCSRs for $B\to \pi\pi$  form
factors in terms of dipion DAs. Hence, the combination of these two independent methods (LCSRs with dipion or $B$-meson
distribution amplitudes) can be used in the future for reciprocal tests of the results. 

Further development of the approach suggested in this paper  is foreseeable  
in several directions. First, the accuracy of the OPE can be improved further, by calculating the perturbative NLO corrections and pinning down the uncertainty in the parameters of the $B$ meson DAs.
Second, the description of the pion vector form factor and probably 
also of the of $B\to \pi\pi$ form factors in the small dipion mass region can be 
implemented in a more model-independent fashion employing  the 
dispersion approach and the Omn\`es representation in the spirit of \Ref{Kang:2013jaa}, that is, with no explicit resonance ansatz. 
Finally, the method can be extended to other $B\to P_1P_2$ form factors with $P_{1,2}=\pi, K,..$ and with various spin-parities and flavor combinations,
for example to $B\to K\pi$ form factors. 

One of the necessary requirements to improve on the accuracy of the observable-rich 
exclusive $B$-decays with unstable mesons in the final states (such as  $B\to \rho \ell \nu_\ell$,  $B\to \rho \pi$ or $B\to K^* \ell^+\ell^-$) is a reliable  and maximally comprehensive 
description of the nonresonant background stemming from excited and continuum
states. Such a description should already begin
at the stage of fitting the data, and the general $B\to \pi\pi$ or $B\to K\pi$ 
form factors, respectively, should serve as a starting point.  
The sum rules considered in this paper provide a useful theoretical tool for 
such purpose. Our analysis is a first step towards a coherent approach to $B$ decays into unstable hadrons.

\section*{Acknowledgments}

We are grateful to G.~Colangelo, S.~Descotes-Genon, L.~Tunstall, D.~van Dyk, Y.M.~Wang and W.F.~Wang for discussions and 
comments. 
This work is supported by the DFG Research Unit FOR 1873 ``Quark Flavour Physics and Effective Theories",
contract No KH 205/2-2.
J.V. acknowledges funding from the Swiss National Science Foundation,
from Explora project FPA2014-61478-EXP,
and from the European Union's Horizon 2020 research and innovation programme under the Marie Sklodowska-Curie grant
agreement No 700525 `NIOBE'.

\newpage

\begin{appendix}

\section{$B$-meson Distribution Amplitudes}\label{app:B-DA}

We use the standard definition of the two-particle $B$-meson DAs $\phi_{\pm}^B(\omega)$ in the momentum representation\cite{Grozin:1996pq,Beneke:2000wa}:
\eqa{
\langle 0 | \bar{d}_{\alpha}(x) [x,0] h_{v\beta}(0) | \bar{B}^0_v \rangle&&
\label{eq:B-WF-3}\\
&&\hspace{-2cm}
= - \frac{i f_B m_B}{4} \int\limits_0^\infty d\omega e^{-i \omega v \cdot x}
\left\{(1+\vsl) \left(\phi_+^B(\omega)+\frac{\phi_-^B(\omega)-\phi_+^B(\omega)}{2v \cdot x} \xsl \right)
\gamma_5 \right\}_{\beta\alpha},\nonumber
}
where $[x,0]$ is the gauge factor and the $B_v$ meson state with four velocity $v$ is defined in HQET. We retain the relativistic normalization $|B(p_B)\rangle = |B_v\rangle$ up to $1/m_b$ corrections; also the $b$ quark field is replaced by the effective field using $b(x)=e^{-im_bvx}h_v(x)$. The variable $\omega$ is the plus component of the light-quark momentum in the $B$ meson. We also use the notation:
\be
\bar{\Phi}_{\pm}^B(\omega)=\int\limits_0^\omega d\tau \left( \phi^B_{+}(\tau)-\phi^B_{-}(\tau)\right).
\label{eq:Phi}
\ee
The four three-particle DAs emerge in the decomposition of the quark-antiquark-gluon matrix element (see \Ref{Khodjamirian:2006st} for details):
\begin{eqnarray}
&&\hspace{-15mm}
\langle 0|\bar{q_2}_\alpha(x) G_{\lambda\rho}(ux)
h_{v\beta}(0)|\bar{B}^0(v)\rangle=
\frac{f_Bm_B}{4}\int\limits_0^\infty d\omega
\int\limits_0^\infty d\xi\,  e^{-i(\omega+u\xi) v\cdot x}
\nonumber \\
&&\hspace{11mm}
\times \Bigg [(1 +\vsl) \Bigg \{ (v_\lambda\gamma_\rho-v_\rho\gamma_\lambda)
\Big(\Psi_A(\omega,\xi)-\Psi_V(\omega,\xi)\Big)
-i\sigma_{\lambda\rho}\Psi_V(\omega,\xi)
\nonumber\\
&&\hspace{11mm}
-\left(\frac{x_\lambda v_\rho-x_\rho v_\lambda}{v\cdot x}\right)X_A(\omega,\xi)
+\left(\frac{x_\lambda \gamma_\rho-x_\rho \gamma_\lambda}{v\cdot x}\right)Y_A(\omega,\xi)\Bigg\}\gamma_5\Bigg]_{\beta\alpha}\,,
\label{eq-B3DAdef}
\end{eqnarray}
where the path-ordered gauge factors in the l.h.s. are omitted for brevity. The DA's $\Psi_{V}$,$\Psi_{A}$, $X_A$ and $Y_A$ depend on $\omega>0$ and $\xi>0$ being, respectively, the plus components of the light-quark and gluon momenta in the $B$ meson.

In the numerical analysis we adopt the popular exponential model \cite{Grozin:1996pq} of the $B$-meson two-particle DAs:
\bea
\phi_+^B(\omega)=\frac{\omega}{\omega_0^2}e^{-\omega/\omega_0};~~~~~\phi_-^B(\omega)=\frac{1}{\omega_0}e^{-\omega/\omega_0},
\label{eq:B-DA-1}
\eea
where the parameter $\omega_0$ is equal to the inverse moment $\lambda_B$, defined as
\be
\frac{1}{\lambda_B}=\int\limits_0 ^\infty d\omega
\frac{\phi_+^B(\omega)}{\omega}\,.
\label{eq:lambB}
\ee
We also use the related models for the three-particle DAs developed in \Ref{Khodjamirian:2006st}:
\begin{eqnarray}
\Psi_A(\omega,\,\xi)& =& \Psi_V(\omega,\,\xi) \,=\,
\dfrac{\lambda_E^2 }{6\omega_0^4}\,\xi^2 e^{-(\omega\,+\,\xi)/\omega_0}\,,
\nonumber\\
X_A(\omega,\,\xi)& = & \dfrac{\lambda_E^2 }{6\omega_0^4}\,
\xi(2\omega-\xi)\,e^{-(\omega\,+\,\xi)/\omega_0}\,,\nonumber\\
Y_A(\omega,\,\xi)& =&  -\dfrac{\lambda_E^2 }{24\omega_0^4}\,
\xi(7\omega_0-13\omega+3\xi)e^{-(\omega\,+\,\xi)/\omega_0}\,.
\label{eq-3partexp}
\end{eqnarray}
where $\lambda^2_E=\frac{3}{2}\omega^2_0$ is adopted.
The three-particle contributions to the sum rules $\Delta V^{BV}$ and $\Delta A_{1,2}^{BV}$ can be found in
\Ref{Khodjamirian:2006st}. However, the expression for $\Delta A_0^{BV}$ has not yet been given
in the literature. In complete analogy with the calculation of \Ref{Khodjamirian:2006st}, we find:
\eqa{
\Delta A_0^{BV}(q^2,s_0^{2\pi},M^2)&=&\int\limits_0^{\sigma_0^{2\pi}}\, d\sigma
\exp\left(\frac{-s(\sigma,q^2)}{M^2}\right)
\times\left( -I^{(A_0)}_1(\sigma) +\frac{I^{(A_0)}_2(\sigma)}{M^2}-
\frac{I^{(A_0)}_3(\sigma)}{2M^4}\right)
\nonumber\\[2mm]
&&
\hspace{-35mm} + \frac{e^{-s_0^{2\pi}/M^2}}{m_B^2}\Bigg\{\eta(\sigma)\Bigg[ I_2^{(A_0)}(\sigma)-\frac12\left( \frac{1}{M^2}
+\frac{1}{m_B^2}\frac{d\eta(\sigma)}{d\sigma}\right)I_3^{(A_0)}(\sigma)
-\frac{\eta(\sigma)}{2m_B^2}\frac{dI_3^{(A_0)}(\sigma)}{d\sigma}\Bigg]\Bigg\}\Bigg
|_{\sigma=\sigma_0}\,, \nonumber\\
}
where
\be
\eta(\sigma)= \left(1-\frac{q^2 }{\bar{\sigma}^2 m_B^2}\right)^{-1}\,,
\nonumber
\ee
and the integrals
over the three-particle DA's multiplying the inverse powers
of the Borel parameter $1/M^{2(n-1)}$ with $n=1,2,3$
are defined as:
\begin{eqnarray}
I^{(A_0)}_n(\sigma)&=& \frac{1}{\bar{\sigma}^n}\int\limits_0^{\sigma m_B} d\omega
\int\limits_{\sigma m_B-\omega}^{\infty}
\frac{d\xi}\xi \Bigg[ C^{(A_0,\Psi A)}_n(\sigma,u,q^2)\Psi_A^{B}(\omega,\xi)
+C^{(A_0,\Psi V)}_n(\sigma,u,q^2)\Psi_V^{B}(\omega,\xi) \non
&+&C^{(A_0,XA)}_n(\sigma,u,q^2)\overline{X}_A^{B}(\omega,\xi)
+C^{(A_0,YA)}_n(\sigma,u,q^2)\overline{Y}_A^{B}(\omega,\xi)
\Bigg]\Bigg|_{u=(\sigma m_B -\omega)/\xi }\quad ,
\label{intn}
\end{eqnarray}
where:
\begin{eqnarray}
\overline{X}_A^{B}(\omega,\xi)=\int\limits_0^\omega d\tau X_A^{B}(\tau,\xi),
\qquad
\overline{Y}_A^{B}(\omega,\xi)=\int\limits_0^\omega d\tau Y_A^{B}(\tau,\xi).
\end{eqnarray}
The non-vanishing coefficients entering Eq.~(\ref{intn}) are given by:
\begin{eqnarray}
C_1^{( A_0^{BV},\Psi A)}&=&-C_1^{( A_0^{BV}, \Psi V)}=\frac{-2 u}{m_B^2\bar{\sigma}}, \non
C_2^{( A_0^{BV},\Psi A)}&=&\frac{(2q^2u+m_B^2\bar{\sigma}(4
\sigma u-2u-3\sigma))}{m_B^2\bar{\sigma}},\non
C_2^{( A_0^{BV},\Psi V)}&=& \frac{(-2q^2u+m_B^2\bar{\sigma}(2u\sigma+2u-3\sigma)}{m_B^2\bar{\sigma}},\non
C_2^{(A_0^{BV},XA)}&=&\frac{m_B(2u-1)(1+\sigma)}{m_B^2\bar{\sigma}},\non
C_2^{(A_0^{BV},YA)}&=&-\frac{2(2u-1)(1+\sigma)}{m_B\bar{\sigma}},\non
C_3^{(A_0^{BV},XA)}&=& \frac{2}{m_B^2\bar{\sigma}}\left[
m_B^3(2u-1)\sigma\bar{\sigma}^2+q^2m_B(1-2u)\sigma\right],\non
C_3^{(A_0^{BV},YA)}&=&-\frac{4}{m_B\bar{\sigma}}\left[
m_B^2(2u-1)\sigma\bar{\sigma}^2+q^2(1-2u)\sigma\right].
\end{eqnarray}

\section{The pion timelike form factor}
\label{app:pi-ff}

We use the results for the pion vector form factor obtained from the measurement of $\tau\to \pi^-\pi^0 \nu_\tau$
decay by Belle Collaboration \cite{Fujikawa:2008ma} and fitted to the combination of three $\rho$-resonances,
for which the Gounaris-Sakurai model \cite{Gounaris:1968mw} was adopted
\bea
F_\pi(s)=\frac{BW^{GS}_\rho(s)+
|\beta|e^{i\phi_\beta}BW^{GS}_{\rho'}(s)+
|\gamma|e^{i\phi_\gamma}BW^{GS}_{\rho''}(s)
}{1+|\beta|e^{i\phi_\beta}+|\gamma|e^{i\phi_\gamma}}\ ,
\label{FpiBelle}
\eea
where
\be
BW^{GS}_R(s)=\frac{m_R^2+m_R\Gamma_Rd}{m_R^2-s+f(s)-i\sqrt{s}\ \Gamma_R(s)}\ ,
\ee
and the functions $f(s)$ and $d$ entering the resonance model are not shown here for the sake of brevity and
can be found e.g., in Eqs.~(14)-(16) of \cite{Fujikawa:2008ma}, so that  $BW^{GS}_R(0)=1$.
(For a simple derivation of the GS model see e.g., \cite{Bruch:2005PY}.) Furthermore,
we use the parameters of the constrained fit ($|F_\pi(0)|=1$) taken from the Table VII of \Ref{Fujikawa:2008ma},
which we reproduce in Table~\ref{TableFpiBelle}. We also note that the masses of resonances and their total
widths are in a good agreement with the averages in \cite{PDG}.

\begin{table}
\centering
\begin{tabular}{@{}cccc@{}}
\toprule
Resonance & $m_R$ (MeV) & $\Gamma_R$ (MeV) & weight factor\\[1mm]
\midrule
$\rho $ & $774.6 \pm 0.2 \pm 0.5$ & $148.1 \pm 0.4\pm 1.7$ & 1.0\\[1mm]
\multirow{2}{*}{$\rho'$} & \multirow{2}{*}{$1446 \pm 7 \pm 28$} & \multirow{2}{*}{$434 \pm 16\pm 60$} &
$|\beta|=0.15 \pm 0.05^{+0.15}_{-0.04}$\\
&&& $\phi_\beta=(202\pm 4 ^{+41}_{-8})\degree$\\[1mm]
\multirow{2}{*}{$\rho''$} & \multirow{2}{*}{$1728 \pm 17 \pm 89$} & \multirow{2}{*}{$164 \pm 21^{+ 80}_{-26}$} &
$|\gamma|=0.037 \pm 0.006^{+0.065}_{-0.009}$\\[1mm]
&&& $\phi_\gamma=(24\pm 9 ^{+118}_{-28})\degree $\\
\bottomrule
\end{tabular}
\caption{\it Numerical parameters for the pion timelike form factor model from \Ref{Fujikawa:2008ma}.}
\label{TableFpiBelle}
\end{table}

\section{Fixing the effective threshold}
\label{app:2pt-s0}

We employ the QCD (SVZ) sum rules \cite{Shifman:1979BX}  for the two-point correlation function:
\be
\Pi_{\mu\nu}(q)=i\int d^4x e^{ikx}\langle 0 | T\{\bar{d}(x)\gamma_\mu u(x),
\bar{u}(0)\gamma_\nu d(0)|0 \rangle = (q_\mu q_\nu-q^2 g_{\mu\nu})\Pi(q^2)\,,
\label{eq:2ptcorr}
\ee
where the lowest two-pion contribution to the hadronic spectral density is written in a  general form, proportional to the square of the pion vector form factor:
\be
\frac{1}{\pi}\im \Pi(s)=\frac{[\beta_\pi(s)]^3}{48 \pi^2} |F_\pi(s)|^2\,.
\label{eq:2ptim}
\ee
It is easy to check that replacing the form factor by the single $\rho$ approximation in the zero-width limit $\Gamma_\rho ^{tot}\to 0 $, brings this expression to the familiar form $\frac{1}{\pi}{\mbox Im} \Pi(s)= f_\rho^2\delta(s-m_\rho^2)$. Furthermore, we substitute in eq.(\ref{eq:2ptim}) the measured form factor squared and calculate numerically the integral over Im$\Pi(s)$ weighted with the Borel exponent:
\be
\Pi^{2\pi}(M^2,s_0^{2\pi})\equiv\frac{1}{\pi}\int\limits_{4m_\pi^2}^{s_0^{2\pi}}ds e^{-s/M^2} \im\Pi(s)= \int\limits_{4m_\pi^2}^{s_0^{2\pi}}ds e^{-s/M^2}\frac{[\beta_\pi(s)]^3}{48 \pi^2} |F_\pi(s)|^2\,.
\ee
The above integral is equated to the Borel-transformed correlation function calculated in QCD and containing the perturbative
loop contribution (to NLO) and the vacuum condensate terms (up to $d=6$):
\be
\Pi^{\rm OPE}(M^2,s_0^{2\pi})= \frac{M^2}{8\pi^2}\big(1-e^{-s_0^{2\pi}/M^2}\big)
\big(1+\frac{\alpha_s}{\pi}\big)+\frac{v_4}{M^2}+\frac{v_6}{2 M^4}\,,
\ee
where
\be
v_4= -\frac{1}{4}f_\pi^2m_\pi^2+ \frac{1}{24}\langle 0 |\frac{\alpha_s}{\pi}G^a_{\mu\nu}G^{a \,\mu\nu} | 0 \rangle\,,
~~
v_6=-\frac{112\pi}{81} \alpha_s\langle 0 |\bar{q}{q}  | 0 \rangle ^2
\label{eq:cond}
\ee
is the compact notation for the contributions from dimension-4 (quark and gluon) and dimension-6 (four-quark)
condensates, respectively. In the above expressions the quark-condensate contribution is related to the pion decay constant $f_\pi=130.4$~MeV
and the input parameters are: $\alpha_s(1~\GeV) =0.47$ \cite{PDG}, $\langle 0 |\bar{q}{q}|0\rangle(1 \GeV)= (-250\pm 10 ~\mbox{MeV})^3 $\cite{Leutwyler:1996qg,PDG} and $\langle 0 |\frac{\alpha_s}{\pi} G^a_{\mu\nu}G^{a \,\mu\nu} | 0 \rangle =0.012^{+ 0.006}_{-0.012}~\GeV^4$ \cite{Ioffe:2002ee}.

Fitting the integral $\Pi^{2\pi}(M^2,s_0^{2\pi})$ to its  QCD sum rule counterpart $\Pi^{\rm OPE}(M^2,s_0^{2\pi})$ we find
the following values depending on the Borel parameter:
\eqa{
s_0^{2\pi}(M^2=1.00\GeV^2) &=&  1.55 \pm 0.04 \GeV^2\nonumber \\
s_0^{2\pi}(M^2=1.25\GeV^2) &=&  1.53 \pm 0.03 \GeV^2\\
s_0^{2\pi}(M^2=1.50\GeV^2) &=&  1.51 \pm 0.02 \GeV^2\nonumber
}
which is close to the duality interval in the original SVZ sum rule \cite{Shifman:1979BX} for the $\rho$ meson.

\end{appendix}

%%%%%%%%%%%%%%%%%%%%%%%%%%%%%%%%


\begin{thebibliography}{99}



\bibitem{Faller:2013dwa}
S.~Faller, T.~Feldmann, A.~Khodjamirian, T.~Mannel and D.~van Dyk,
``Disentangling the Decay Observables in $B^- \to \pi^+\pi^-\ell^-\bar\nu_\ell$,''
Phys.\ Rev.\ D {\bf 89}, 014015 (2014)
%doi:10.1103/PhysRevD.89.014015
[arXiv:1310.6660 [hep-ph]].
%%CITATION = doi:10.1103/PhysRevD.89.014015;%%

\bibitem{Kang:2013jaa}
X.~W.~Kang, B.~Kubis, C.~Hanhart and U.~G.~Mei\ss ner,
``$B_{l4}$ decays and the extraction of $|V_{ub}|$,''
Phys.\ Rev.\ D {\bf 89}, 053015 (2014),
%doi:10.1103/PhysRevD.89.053015
[arXiv:1312.1193 [hep-ph]].
%%CITATION = doi:10.1103/PhysRevD.89.053015;%%

\bibitem{Sibidanov:2013rkk}
{\bf Belle} collaboration,
``Study of Exclusive $B \to X_u \ell \nu$ Decays and Extraction of $|V_{ub}|$ using Full Reconstruction Tagging at the Belle Experiment,''
Phys.\ Rev.\ D {\bf 88}, no. 3, 032005 (2013)
%doi:10.1103/PhysRevD.88.032005
[arXiv:1306.2781 [hep-ex]].
%%CITATION = doi:10.1103/PhysRevD.88.032005;%%


\bibitem{Aaij:2014lba}
{\bf LHCb} collaboration,
``Study of the rare $B_s^0$ and $B^0$ decays into the $\pi^+\pi^-\mu^+\mu^-$ final state,''
Phys.\ Lett.\ B {\bf 743}, 46 (2015)
%  doi:10.1016/j.physletb.2015.02.010
[arXiv:1412.6433 [hep-ex]].
%%CITATION = doi:10.1016/j.physletb.2015.02.010;%%


\bibitem{Krankl:2015fha}
S.~Kr\"ankl, T.~Mannel and J.~Virto,
``Three-Body Non-Leptonic B Decays and QCD Factorization,''
Nucl.\ Phys.\ B {\bf 899}, 247 (2015),
%  doi:10.1016/j.nuclphysb.2015.08.004
[arXiv:1505.04111 [hep-ph]].
%%CITATION = doi:10.1016/j.nuclphysb.2015.08.004;%%

\bibitem{1609.07430} 
  J.~Virto,
  ``Charmless Non-Leptonic Multi-Body B decays,''
  PoS FPCP {\bf 2016}, 007 (2016),
  arXiv:1609.07430 [hep-ph].


\bibitem{BoerFeldmannDyk} 
  P.~B\"oer, T.~Feldmann and D.~van Dyk,
  ``QCD Factorization Theorem for $B \to \pi\pi\ell\nu$ Decays at Large Dipion Masses,''
  arXiv:1608.07127 [hep-ph]



\bibitem{Hambrock:2015aor}
C.~Hambrock and A.~Khodjamirian,
``Form factors in $\bar B^0 \to \pi\pi\ell\bar\nu_\ell$ from QCD light-cone sum rules,''
Nucl.\ Phys.\ B {\bf 905}, 373 (2016),
%doi:10.1016/j.nuclphysb.2016.02.035
[arXiv:1511.02509 [hep-ph]].
%%CITATION = doi:10.1016/j.nuclphysb.2016.02.035;%%

\bibitem{Diehl:1998dk}
M.~Diehl, T.~Gousset, B.~Pire and O.~Teryaev,
``Probing partonic structure in $\gamma^* \gamma \to \pi \pi$ near threshold,''
Phys.\ Rev.\ Lett.\  {\bf 81} (1998) 1782,
%doi:10.1103/PhysRevLett.81.1782
[hep-ph/9805380].
%%CITATION = doi:10.1103/PhysRevLett.81.1782;%%

\bibitem{Polyakov:1999ZE}
M.V.~Polyakov,
''Hard exclusive electroproduction of two pions and their resonances,''
Nucl.\ Phys.\ B {\bf 555}, 231 (1999),
%doi:10.1016/S0550-3213(99)00314-4
[arXiv:hep-ph/9809483].

\bibitem{Khodjamirian:2005ea}
A.~Khodjamirian, T.~Mannel and N.~Offen,
``$B$-meson distribution amplitude from the $B \to \pi$ form-factor,''
Phys.\ Lett.\ B {\bf 620}, 52 (2005),
%  doi:10.1016/j.physletb.2005.06.021
[hep-ph/0504091].
%%CITATION = doi:10.1016/j.physletb.2005.06.021;%%

\bibitem{Khodjamirian:2006st}
A.~Khodjamirian, T.~Mannel and N.~Offen,
``Form-factors from light-cone sum rules with B-meson distribution amplitudes,''
Phys.\ Rev.\ D {\bf 75}, 054013 (2007),
%doi:10.1103/PhysRevD.75.054013
[hep-ph/0611193].

\bibitem{DeFazio:2005dx}
F.~De Fazio, T.~Feldmann and T.~Hurth,
``Light-cone sum rules in soft-collinear effective theory,''
Nucl.\ Phys.\ B {\bf 733}, 1 (2006),
[arXiv:hep-ph/0504088].

\bibitem{Eidelman:2003uh}
S.~Eidelman and L.~Lukaszuk,
``Pion form-factor phase, $\pi \pi$ elasticity and new $e^+ e^-$ data,''
Phys.\ Lett.\ B {\bf 582}, 27 (2004),
%doi:10.1016/j.physletb.2003.12.030
[hep-ph/0311366].
%%CITATION = doi:10.1016/j.physletb.2003.12.030;%%

\bibitem{Fujikawa:2008ma}
{\bf Belle} collaboration,
``High-Statistics Study of the $\tau \to \pi^- \pi^0 \nu_\tau$ Decay,''
Phys.\ Rev.\ D {\bf 78} (2008) 072006,
%doi:10.1103/PhysRevD.78.072006
[arXiv:0805.3773 [hep-ex]].


\bibitem{Aubert:2009AD}
{\bf BaBar} collaboration,
``Precise measurement of the $e^+ e^- \to \pi^+ \pi^- (\gamma)$ cross section with the Initial State Radiation method at BABAR,''
Phys.\ Rev.\ Lett.\  {\bf 103}, 231801 (2009),
[arXiv:0908.3589 [hep-ex]].
%%CITATION = doi:10.1103/PhysRevLett.103.231801%%



\bibitem{PDG}
{\bf PDG} collaboration,
  ``Review of Particle Physics,''
  Chin.\ Phys.\ C {\bf 40}, no. 10, 100001 (2016).
  %doi:10.1088/1674-1137/40/10/100001
  %%CITATION = doi:10.1088/1674-1137/40/10/100001;%%

\bibitem{Bruch:2005PY}
C.~Bruch, A.~Khodjamirian and J.~H.~K\"uhn,
``Modeling the pion and kaon form factors in the timelike region,''
Eur.\ Phys.\ J.\ C {\bf 39}, 41 (2005),
[arXiv:hep-ph/0409080].
%%CITATION = doi:10.1140/epjc/s2004-02064-3;%%

\bibitem{Czyz:2010hj}
H.~Czyz, A.~Grzelinska and J.~H.~K\"uhn,
``Narrow resonances studies with the radiative return method,''
Phys.\ Rev.\ D {\bf 81}, 094014 (2010),
%doi:10.1103/PhysRevD.81.094014
[arXiv:1002.0279 [hep-ph]].
%%CITATION = doi:10.1103/PhysRevD.81.094014;%%

\bibitem{Hanhart:2012WI}
C.~Hanhart,
``A New Parameterization for the Pion Vector Form Factor,''
Phys.\ Lett.\ B {\bf 715}, 170 (2012),
[arXiv:1203.6839 [hep-ph]].

\bibitem{Achasov:2011RA}
N.~N.~Achasov and A.~A.~Kozhevnikov,
``Electromagnetic form factor of pion in the field theory inspired  approach,''
Phys.\ Rev.\ D {\bf 83}, 113005 (2011),
Erratum: [Phys.\ Rev.\ D {\bf 85}, 019901  (2012)],
[arXiv:1104.4225 [hep-ph]].
%%CITATION = doi:10.1103/PhysRevD.83.113005, 10.1103/PhysRevD.85.019901%%

\bibitem{Shekhovtsova:2012RA}
O.~Shekhovtsova, T.~Przedzinski, P.~Roig and Z.~Was,
``Resonance chiral Lagrangian currents and $\tau$ decay Monte Carlo,''
Phys.\ Rev.\ D {\bf 86}, 113008 (2012),
[arXiv:1203.3955 [hep-ph]].


\bibitem{Dominguez:2001zu}
C.~A.~Dominguez,
``Pion form-factor in large $N_c$ QCD,''
Phys.\ Lett.\ B {\bf 512}, 331 (2001),
%doi:10.1016/S0370-2693(01)00576-7
[hep-ph/0102190].
%%CITATION = doi:10.1016/S0370-2693(01)00576-7;%%

\bibitem{Gounaris:1968mw}
G.~J.~Gounaris and J.~J.~Sakurai,
``Finite width corrections to the vector meson dominance prediction for $\rho \to e^+ e^-$,''
Phys.\ Rev.\ Lett.\  {\bf 21}, 244 (1968),
%doi:10.1103/PhysRevLett.21.244
%%CITATION = doi:10.1103/PhysRevLett.21.244;%%

\bibitem{Kuhn:1990ad}
J.~H.~K\"uhn and A.~Santamaria,
``Tau decays to pions,''
Z.\ Phys.\ C {\bf 48} (1990) 445.
%doi:10.1007/BF01572024
%%CITATION = doi:10.1007/BF01572024;%%

\bibitem{BCL}
C.~Bourrely, I.~Caprini and L.~Lellouch,
``Model-independent description of $B \to \pi \ell \nu$ decays and a determination of $|V_{ub}|$,''
Phys.\ Rev.\ D {\bf 79}, 013008 (2009),
Erratum: [Phys.\ Rev.\ D {\bf 82}, 099902 (2010)],
%doi:10.1103/PhysRevD.82.099902, 10.1103/PhysRevD.79.013008
[arXiv:0807.2722 [hep-ph]].
%%CITATION = doi:10.1103/PhysRevD.82.099902, 10.1103/PhysRevD.79.013008;%%

\bibitem{KMPW}
A.~Khodjamirian, T.~Mannel, A.~A.~Pivovarov and Y.-M.~Wang,
``Charm-loop effect in $B \to K^{(*)} \ell^{+} \ell^{-}$ and $B\to K^*\gamma$,''
JHEP {\bf 1009}, 089 (2010),
%doi:10.1007/JHEP09(2010)089
[arXiv:1006.4945 [hep-ph]].
%%CITATION = doi:10.1007/JHEP09(2010)089;%%

\bibitem{Braun:2003wx}
V.~M.~Braun, D.~Y.~Ivanov and G.~P.~Korchemsky,
``The B-meson distribution amplitude in QCD,''
Phys.\ Rev.\ D {\bf 69} (2004) 034014,
[arXiv:hep-ph/0309330].
%%CITATION = HEP-PH 0309330;%%

\bibitem{Beneke:2015wfa}
M.~Beneke,
``Soft-collinear factorization in $B$ decays,''
Nucl.\ Part.\ Phys.\ Proc.\  {\bf 261-262}, 311 (2015),
%doi:10.1016/j.nuclphysbps.2015.03.021
[arXiv:1501.07374 [hep-ph]].
%%CITATION = doi:10.1016/j.nuclphysbps.2015.03.021;%%


\bibitem{Beneke:2009ek} 
  M.~Beneke, T.~Huber and X.~Q.~Li,
  ``NNLO vertex corrections to non-leptonic B decays: Tree amplitudes,''
  Nucl.\ Phys.\ B {\bf 832}, 109 (2010)
  %doi:10.1016/j.nuclphysb.2010.02.002
  [arXiv:0911.3655 [hep-ph]].
  %%CITATION = doi:10.1016/j.nuclphysb.2010.02.002;%%

\bibitem{Heller:2015vvm}
A.~Heller {\it et al.} [Belle Collaboration],
``Search for $B^+ \to \ell^+ \nu_\ell\gamma $ decays with hadronic tagging using the full Belle data sample,''
Phys.\ Rev.\ D {\bf 91} (2015) no.11,  112009,
%doi:10.1103/PhysRevD.91.112009
[arXiv:1504.05831 [hep-ex]].
%%CITATION = doi:10.1103/PhysRevD.91.112009;%%

\bibitem{Beneke:2011nf}
M.~Beneke and J.~Rohrwild,
``$B$-meson distribution amplitude from $B \to \gamma \ell \nu$,''
Eur.\ Phys.\ J.\ C {\bf 71}, 1818 (2011),
%  doi:10.1140/epjc/s10052-011-1818-8
[arXiv:1110.3228 [hep-ph]].
%%CITATION = doi:10.1140/epjc/s10052-011-1818-8;%%

\bibitem{Braun:2012kp}
V.~M.~Braun and A.~Khodjamirian,
``Soft contribution to $B\to \gamma \ell \nu_\ell$ and the $B$-meson distribution amplitude,''
Phys.\ Lett.\ B {\bf 718}, 1014 (2013),
% doi:10.1016/j.physletb.2012.11.047
[arXiv:1210.4453 [hep-ph]].
%%CITATION = doi:10.1016/j.physletb.2012.11.047;%%

\bibitem{Wang:2015VGV}
Y.~M.~Wang and Y.~L.~Shen,
``QCD corrections to $B\to \pi$ form factors from light-cone sum rules,''
Nucl.\ Phys.\ B {\bf 898}, 563 (2015),
[arXiv:1506.00667 [hep-ph]].
%%CITATION = doi: 10.1016/j.nuclphysb.2015.07.016%%

\bibitem{Khodjamirian:2011ub}
A.~Khodjamirian, T.~Mannel, N.~Offen and Y.-M.~Wang,
``$B \to \pi \ell \nu_l$ Width and $|V_{ub}|$ from QCD Light-Cone Sum Rules,''
Phys.\ Rev.\ D {\bf 83}, 094031 (2011),
%doi:10.1103/PhysRevD.83.094031
[arXiv:1103.2655 [hep-ph]].
%%CITATION = doi:10.1103/PhysRevD.83.094031;%%


%\cite{Gelhausen:2013wia}
\bibitem{Gelhausen:2013wia}
  P.~Gelhausen, A.~Khodjamirian, A.~A.~Pivovarov and D.~Rosenthal,
  ``Decay constants of heavy-light vector mesons from QCD sum rules,''
  Phys.\ Rev.\ D {\bf 88} (2013) 014015,
   Errata: [Phys.\ Rev.\ D {\bf 89} (2014) 099901, Phys.\ Rev.\ D {\bf 91} (2015) 099901]
  %doi:10.1103/PhysRevD.88.014015, 10.1103/PhysRevD.91.099901, 10.1103/PhysRevD.89.099901
  [arXiv:1305.5432 [hep-ph]].
  %%CITATION = doi:10.1103/PhysRevD.88.014015, 10.1103/PhysRevD.91.099901, 10.1103/PhysRevD.89.099901;%%
  %52 citations counted in INSPIRE as of 17 Aug 2016


\bibitem{FLAG} 
  {\bf FLAG} collaboration,
  ``Review of lattice results concerning low-energy particle physics,''
  %doi:10.3204/PUBDB-2016-03637
  arXiv:1607.00299 [hep-lat].
  %%CITATION = doi:10.3204/PUBDB-2016-03637;%%

\bibitem{Shifman:1979BX}
M.~A.~Shifman, A.~I.~Vainshtein and V.~I.~Zakharov,
``QCD and Resonance Physics. Theoretical Foundations,''
Nucl.\ Phys.\ B {\bf 147}, 385 (1979).
%%CITATION = doi:10.1016/0550-3213(79)90022-1%%


\bibitem{BSZ}
A.~Bharucha, D.~M.~Straub and R.~Zwicky,
``$B\to V\ell^+\ell^-$ in the Standard Model from Light-Cone Sum Rules,''
JHEP {\bf 1608}, 098 (2016), [arXiv:1503.05534 [hep-ph]].
%%CITATION = ARXIV:1503.05534;%%

\bibitem{Ball:2005RG}
P.~Ball and R.~Zwicky,
``$B_{d,s} \to \rho, \omega, K^*, \phi$ decay form-factors from light-cone sum rules revisited,''
Phys.\ Rev.\ D {\bf 71}, 014029 (2005),
[arXiv:hep-ph/0412079].\\
%%CITATION = doi:10.1103/PhysRevD.71.014029%%
%

\bibitem{Grozin:1996pq}
A.~G.~Grozin and M.~Neubert,
``Asymptotics of heavy-meson form factors,''
Phys.\ Rev.\ D {\bf 55} (1997) 272,
[arXiv:hep-ph/9607366].
%%CITATION = HEP-PH 9607366;%%


\bibitem{Beneke:2000wa}
M.~Beneke and T.~Feldmann,
``Symmetry breaking corrections to heavy to light B meson form-factors at large recoil,''
Nucl.\ Phys.\ B {\bf 592} (2001) 3,
%doi:10.1016/S0550-3213(00)00585-X
[hep-ph/0008255].
%%CITATION = doi:10.1016/S0550-3213(00)00585-X;%%

\bibitem{Leutwyler:1996qg}
H.~Leutwyler,
``The ratios of the light quark masses,''
Phys.\ Lett.\ B {\bf 378} (1996) 313,
%doi:10.1016/0370-2693(96)00386-3
[hep-ph/9602366].
%%CITATION = doi:10.1016/0370-2693(96)00386-3;%%

\bibitem{Ioffe:2002ee}
B.~L.~Ioffe,
``Condensates in quantum chromodynamics,''
Phys.\ Atom.\ Nucl.\  {\bf 66} (2003) 30,
Yad.\ Fiz.\  {\bf 66} (2003) 32,
% doi:10.1134/1.1540654
[hep-ph/0207191].
%%CITATION = doi:10.1134/1.1540654;%%




\end{thebibliography}
\end{document}